\documentclass[twoside,11pt]{article}

\usepackage{blindtext}

\usepackage[preprint]{jmlr2e}
\usepackage{mathtools}
\usepackage{multirow}
\usepackage{bbm} 
\usepackage{algorithm}
\usepackage{algpseudocode}
\usepackage{amsmath,amssymb}
\usepackage{xcolor}

\newcommand{\R}{\mathbb{R}}

\newcommand{\E}{\mathbb{E}}

\newcommand{\sX}{\mathcal{X}}
\newcommand{\sY}{\mathcal{Y}}
\newcommand{\sZ}{\mathcal{Z}}
\newcommand{\sP}{\mathcal{P}}

\def\sO {\mathcal{O}}

\renewcommand{\d}{\mathrm{d}} 

\newcommand{\Id}{\mathrm{Id}}
\newcommand{\KL}{\mathrm{KL}}
\newcommand{\rTV}{\right\|_{TV}}
\newcommand{\lTV}{\left\|}
\newcommand{\gap}{\mathrm{Gap}}
\newcommand{\simiid}{\overset{\text{i.i.d.}}{\sim}}

\newcommand{\PDAtilde}{\tilde{P}_\mathrm{DA}}
\newcommand{\PDA}{P_\mathrm{DA}}
\newcommand{\PDAmod}{P_\mathrm{DA, mod}}
\newcommand{\PMG}{P_\mathrm{MG}}
\newcommand{\PGS}{P_\mathrm{CG}}
\newcommand{\PGSi}{P_{\mathrm{CG}, i}}

\newcommand{\tmix}{\tau_{\mathrm{mix}}}
\newcommand{\tmixchi}{\tau_{\mathrm{mix}, 2}}

\newtheorem{asmp}{Assumption}

\firstpageno{1}

\usepackage{lastpage}
\jmlrheading{27}{2026}{1-\pageref{LastPage}}{9/25; Revised
5/26}{6/26}{25-2192}{Filippo Ascolani and Giacomo Zanella}
\ShortHeadings{Gibbs samplers for high-dimensional probit regression}{Ascolani and Zanella}
\begin{document}

\title{Mixing times of data-augmentation Gibbs samplers for high-dimensional probit regression}

\author{\name Filippo Ascolani  \email filippo.ascolani@duke.edu \\
       \addr Department of Statistical Science, \\
      Duke University\\
       Durham, NC 27708, USA
       \AND
       \name Giacomo Zanella \email giacomo.zanella@unibocconi.it \\
       \addr Department of Decision Sciences and BIDSA,\\
       Bocconi University\\
       Milano, 20136, Italy}

\editor{Anthony Lee}

\maketitle

\begin{abstract}
We investigate the convergence properties of popular data-augmentation samplers for Baye\-sian probit regression. Leveraging recent results on Gibbs samplers for log-concave targets, we provide simple and explicit non-asymptotic bounds on the associated mixing times (in Kullback-Leibler divergence). The bounds depend explicitly on the design matrix and the prior precision, while they hold uniformly over the vector of responses. We specialize the results for different regimes of statistical interest, when both the number of data points $n$ and parameters $p$ are large: in particular we identify scenarios where the mixing times remain bounded as $n,p\to\infty$, and ones where they do not. The results are shown to be tight (in the worst case with respect to the responses) and provide guidance on choices of prior distributions that provably lead to fast mixing. An empirical analysis based on coupling techniques suggests that the bounds are effective in predicting practically observed behaviours.
\end{abstract}


\section{Introduction}\label{sec:intro}

\subsection{The model}\label{sec:probit_model}

Probit regression is a popular methodology when the relationship between binary data $y_i \in \{0, 1\}$ and a set of predictors is of interest. It is an instance of generalized linear model \citep{glm1989} and its usual Bayesian formulation reads
\begin{align}\label{eq:model_orig}
y_i&\mid \beta \sim \text{Bernoulli}(\Phi(x_i^T\beta)), \quad  \beta \sim N(m,Q_0^{-1}),&i = 1, \dots, n
\end{align}
where $x_i\in\R^p$ is the $i$-th row of a design matrix $X \in \R^{n \times p}$, $\Phi(\cdot)$ is the cumulative distribution function of a standard Gaussian random variable and $N(m, Q_0^{-1})$ denotes the multivariate normal distribution with mean $m$ and precision matrix $Q_0$.
Given data $y = (y_1, \dots, y_n)\in\{0,1\}^n$, the posterior distribution of $\beta$ has density
\begin{equation}\label{eq:model}
\pi(\beta) \, \propto\, N(\beta \mid m,Q_0^{-1})\prod_{i = 1}^n\Phi(x_i^T\beta)^{y_i}(1-\Phi(x_i^T\beta))^{1-y_i},
\end{equation}
where $N(\beta \mid m, Q_0^{-1})$ denotes the density of $N(m, Q_0^{-1})$ at $\beta$. 
Several strategies have been developed to approximate this distribution, ranging from exact rejection sampling \citep{botev2017normal,durante2019conjugate}, variational inference \citep{chopin2017leave,fasano2022scalable} and sampling with Markov chain Monte Carlo (MCMC) techniques \citep{albert1993bayesian, held2006bayesian}, which is the focus of this article. A popular class of MCMC methods for $\pi(\beta)$ relies on a data augmentation scheme based on re-writing model \eqref{eq:model_orig} as
\begin{equation}\label{eq:probit_DA}
\begin{aligned}
&y_i
= \mathbbm{1}(z_i > 0)&i=1,\dots,n,\\
&z|\beta \sim N(X\beta,I_n), \quad \beta \sim N(m,Q_0^{-1}),
\end{aligned}
\end{equation}
where $z=(z_1,\dots,z_n)^T\in\R^n$, $I_n$ denotes the $n\times n$ identity matrix and $\mathbbm{1}$ denotes the indicator function. 
The joint posterior density of $z$ and $\beta$ is
\begin{equation}\label{eq:posterior}
\pi(z, \beta) \, \propto \, N(\beta \mid m,Q_0^{-1})N(z \mid X\beta,I_n)\prod_{i = 1}^n\mathbbm{1}\left(y_i = g(z_i)\right),
\end{equation}
where $g(z_i) = \mathbbm{1}(z_i > 0)$, and its marginal density over $\beta$ coincides with \eqref{eq:model}.

\subsection{The algorithms}\label{sec:algorithms}
We consider two popular Gibbs Sampling schemes used to draw samples from $\pi(z,\beta)$. 

\subsubsection{Data Augmentation (DA)} 
First, we consider the two-block deterministic-scan Gibbs Sampler originally proposed in \cite{albert1993bayesian}, which alternates the update of $z$ from 
$$\pi(z \mid \beta) \, \propto \, N(z \mid X\beta, I_n)\prod_{i = 1}^n\mathbbm{1}(y_i = g(z_i))$$
and $\beta$ from 
\begin{equation}\label{eq:posterior_beta}
\pi(\beta \mid z) = N\left(\beta \mid (X^TX+Q_0)^{-1}(Q_0m+X^Tz) , (X^TX+Q_0)^{-1}\right)\,.
\end{equation} 
Equivalently, its Markov kernel $\PDA$ is defined as the composition of two kernels, $\PDA= P_\beta P_z$,
with
\begin{align}\label{eq:AC_algorithm}
P_z((z,\beta), (\d z', \d \beta')) = \delta_{\beta}(\d \beta')\pi(\d z' \mid \beta)
\quad\hbox{and}\quad P_\beta((z,\beta), (\d z', \d \beta'))= \delta_{z}(\d z')\pi(\d \beta' \mid z)
\end{align} 
for $z\in\R^n$ and $\beta\in\R^p$.
The pseudocode for $\PDA$ is given in Algorithm \ref{alg:PDA}.
Note that the conditional distribution $\pi(z\mid\beta)$ factorizes across the $n$ components of $z$, so that $P_z$ entails sampling $n$ independent truncated normal random variables. Thus, both $P_z$ and $P_\beta$ can be implemented in closed form, which is the main computational advantage of the latent variable representation in \eqref{eq:probit_DA}.

\begin{algorithm}[htbp]
\begin{algorithmic}
\State Initialize $\beta^{(0)}$. 
\For{$t=1,2,\dots$}
    \State Sample $z^{(t)}_i \sim \pi(z_i \mid \beta^{(t-1)})\, \propto \, N(z_i \mid x_i^T\beta^{(t-1)}, 1)\mathbbm{1}(y_i = g(z_i))$ independently for $i = 1, \dots, n$.
    \State Sample $\beta^{(t)} \sim \pi(\beta \mid z^{(t)})$ with $\pi(\beta \mid z)$ as in \eqref{eq:posterior_beta}.
   \EndFor
\end{algorithmic}
\caption{(Data Augmentation Gibbs sampler $\PDA$)
\label{alg:PDA}}
\end{algorithm}

\subsubsection{Collapsed Gibbs (CG)} 

Second, we consider the $n$-blocks random scan Gibbs sampler on $\sX = \R^n$ targeting 
\begin{equation}\label{eq:posterior_marginal_zeta}
\pi(z) 
=
\int_{\R^p} \pi(z,\beta)d\beta
\; \propto \; 
N(z \mid Xm, I_n + XQ_0^{-1}X^T)\prod_{i = 1}^n\mathbbm{1}(y_i = g(z_i)),
\end{equation}
which we refer to as \emph{Collapsed Gibbs} (CG) sampler.
Its Markov kernel $\PGS$ is defined as
\begin{equation}\label{eq:GS_algorithm}
\PGS(z, \d z') = \frac{1}{n}\sum_{i = 1}^n\PGSi(z, \d z'), \quad\hbox{ with } \PGSi(z, \d z') = \delta_{z_{-i}}(\d z'_{-i})\pi(\d z'_i \mid z_{-i}),
\end{equation}
where $\pi(\d z_i \mid z_{-i})$ is the conditional distribution with density
\begin{equation}\label{eq:post_marginal_z}
\pi(z_i \mid z_{-i}) \, \propto \, N\left(z_i \mid (1-h_i)^{-1}x_i^TVX^T(z-Q_0m)-h_i(1-h_i)^{-1}z_i, (1-h_i)^{-1}\right)\mathbbm{1}(y_i = g(z_i)),
\end{equation}
with $V = (X^TX+Q_0)^{-1}$ and $h_i = x_i^TVx_i$. See Section $2$ in \cite{held2006bayesian} for a derivation of $\pi(z)$ and its full conditionals.
The pseudocode for $\PGS$ is given in Algorithm \ref{alg:PGS}.
\begin{algorithm}[htbp]
\begin{algorithmic}
\State Initialize $z^{(0)}\in\R^n$. 
\For{$t \geq 1$}
    \State Sample $I$ uniformly at random from $\{1, \dots, n\}$.
    \State Given $I = i$, sample $z^{(t)}_i \sim \pi(z_i \mid z_{-i}^{(t-1)})$, with $\pi(z_i \mid z_{-i})$ as in \eqref{eq:post_marginal_z}.
   \EndFor
\end{algorithmic}
\caption{(Collapsed Gibbs sampler $\PGS$)
\label{alg:PGS}}
\end{algorithm}
The kernel $\PGS$ can be used to sample from $\pi(z)$ and, given a sample from $\pi(z)$, one can obtain samples from $\pi(z,\beta)$ by drawing $\beta$ from $\pi(\beta \mid z)$ defined in \eqref{eq:posterior_beta}.
The deterministic scan version of $\PGS$ was originally considered in \cite{held2006bayesian}. Here we consider the random scan version for theoretical convenience, since the latter is easier to analyse in our context.


\section{Main results}\label{sec:main_results}

\subsection{$\KL$-mixing times}
In this paper we measure distance to stationarity
using the Kullback-Leibler (KL) divergence.
For every $\mu, \nu\in \sP(\sX)$, where $\sP(\sX)$ denotes the set of probability measures over $\sX$, let
\[
\KL(\mu, \nu) = \int_{\sX}\log\left(\frac{\d\mu}{\d\nu}(x) \right)\mu(\d x),
\]
where $\frac{\d\mu}{\d\nu}$ denotes the Radon-Nikodym derivative between $\mu$ and $\nu$.  
Given a $\pi$-invariant Markov kernel $P$ and a starting distribution $\mu \in \sP(\sX)$, we define the mixing times with respect to $\KL$ as
\begin{equation}\label{eq:mixing_definition}
    \tmix(\epsilon,\mu, P):=\inf\{t\geq 1\,:\,\KL( \mu P^t , \pi ) 
\leq 
\epsilon\}, \qquad \epsilon\in[0,\infty),
\end{equation}
where $\mu P^t(A) = \int_{\sX}P^t(x,A)\mu(\d x)$ for every $A\subseteq \sX$. 
In words, $\tmix(\epsilon,\mu, P)$ is the number of iterations needed for the chain to be $\epsilon$-close in $\KL$ to its stationary distribution $\pi$, when starting from $\mu$.

\subsection{Mixing time bounds}
The goal of this work is to quantify the computational effort required by $\PDA$ and $\PGS$ to produce approximate samples from $\pi(z, \beta)$. The key task in doing so is to upper bound their mixing times.  
Since the cost per iteration of $\PDA$ is $n$ times larger than that of $\PGS$ (see Section A of the Supplementary Material for details), we will compare a single iteration of $\PDA$ to $n$ iterations of $\PGS$. In other words, we express results in terms of $\tmix(\epsilon,\mu, \PDA)$ and $\tmix(\epsilon,\mu_1, \PGS^n)$, where 
$\mu \in \sP(\R^n\times \R^p)$ and $\mu_1 \in \sP(\R^n)$ denotes the first marginal of $\mu$. The next theorem provides an explicit upper bound to these two quantities.
\begin{theorem}\label{theo:mixing_times}
For every $\mu \in \sP(\R^n\times \R^p)$ and $\epsilon>0$, we have
\begin{align}\label{eq:bound_mix_time_GS}
\tmix(\epsilon,\mu, \PDA) &\leq (2+\lambda_{max}(XQ_0^{-1}X^T))\log\left(\frac{\KL(\mu, \pi)}{\epsilon}\right)\,,
\end{align}
and
\begin{align}\label{eq:bound_mix_time_PCG}
\tmix(\epsilon,\mu_1, \PGS^n) &\leq \frac{1+\lambda_{max}(XQ_0^{-1}X^T)}{1+\lambda_{min}(XQ_0^{-1}X^T)}\log\left(\frac{\KL(\mu, \pi)}{\epsilon}\right)\,,
\end{align}
where $\lambda_{min}$ and $\lambda_{max}$ denote the minimum and maximum (modulus) eigenvalues of the given matrix.
\end{theorem}
\begin{proof}
The proof is postponed to Section \ref{sec:proof_main} of the Supplementary Material.
\end{proof}
In Section \ref{sec:starting_distribution} we 
provide a so-called feasible starting distribution $\mu$ such that $\log(\KL(\mu, \pi))$ is at most of order $\log\big(n + n\log\big(n\lambda_{max}(XQ_0^{-1}X^T)\big)\big)$, see Proposition \ref{prop:starting_distribution}. Thus, by Theorem \ref{theo:mixing_times}, the mixing times of both $\PDA$ and $\PGS^n$ are upper bounded by $\lambda_{max}(XQ_0^{-1}X^T)$, up to constants and logarithmic terms. 

\begin{remark}
The bounds in Theorem \ref{theo:mixing_times} hold for every fixed $X$ and $y$. 
The right-hand side is independent of $y$, which means that it considers the worst-case with respect to the data $y$. This will be important in interpreting the results later on.
\end{remark}

\begin{remark}\label{rmk:results}
The statements in Theorem \ref{theo:mixing_times} are actually a consequence of a stronger result. For example in the case of $\PGS$ we upper bound the one-step entropy contraction as
\begin{equation}\label{eq:one_step_contraction_PCG}
\frac{\KL(\mu_1 \PGS, \pi_1)}{\KL(\mu_1, \pi_1)} \leq 1-\frac{1}{n}\left[\frac{1+ \lambda_{min}(XQ_0^{-1}X^T)}{1+ \lambda_{max}(XQ_0^{-1}X^T)}\right],
\end{equation}
for every $\mu_1 \in \sP(\R^n)$ such that $\KL(\mu_1, \pi_1) < \infty$. This implies \eqref{eq:bound_mix_time_PCG}, see \eqref{eq:mix_rho_EC} and Theorem \ref{theo:contraction_GS} in Appendix \ref{sec:proofs_appendix}. The case of $\PDA$ is more complex, due to the lack of reversibility: we provide a general framework to study entropy contraction for data augmentation schemes in Section \ref{sec:theory}.
Bounds as in \eqref{eq:one_step_contraction_PCG} allow to study mixing times in other metrics. For example, denoting with
\[
\chi^2(\mu, \nu) = \int_{\sX}\left(\frac{\d\nu}{\d\mu}(x)-1 \right)^2\nu(\d x), \qquad \mu, \nu \in \sP(\sX)
\]
the chi-square divergence and with $\tmixchi$ the mixing times obtained by replacing $\KL(\mu P^t, \pi)$ in \eqref{eq:mixing_definition} with $\chi^2(\mu P^t, \pi)$, we have
\begin{align*}
\tmixchi(\epsilon,\mu, \PDA) &\leq (3+2\lambda_{max}(XQ_0^{-1}X^T))\log\left(\frac{\chi^2(\mu, \pi)}{\epsilon}\right)
\end{align*}
and
\begin{align*}
\tmixchi(\epsilon,\mu_1, \PGS^n) &\leq 2\,\frac{1+\lambda_{max}(XQ_0^{-1}X^T)}{1+\lambda_{min}(XQ_0^{-1}X^T)}\log\left(\frac{\chi^2(\mu, \pi)}{\epsilon}\right)\,.
\end{align*}
See Section \ref{sec:mixing_times_chi} and Theorem \ref{theo:mixing_times_chi} in the Supplementary Material for details and a formal proof. Since $\chi^2(\mu, \nu) \geq \KL(\mu, \nu)$ the latter inequalities are not implied by \eqref{eq:bound_mix_time_GS} and \eqref{eq:bound_mix_time_PCG} alone: moreover at this level of generality they are tight, see Proposition \ref{prop:lower_bound_KL} for a matching lower bound in a special case.
\end{remark}

\begin{remark}\label{rmk:model_variations}
While in the main body of the paper we focus on probit regression for simplicity, $\PDA$ and $\PGS$ can also be used to sample from other popular models that can be expressed as partially or fully discretized Gaussian linear regression, such as multinomial probit and tobit regressions, see e.g.\ \cite{anceschi2023bayesian} for a recent review.  
In Section \ref{sec:tobit} in the supplementary material we illustrate extensions of Theorem \ref{theo:mixing_times} to those cases.
\end{remark}


\subsubsection{Proof technique} The main idea underlying the proofs is to recognize that the target distributions of both $\PDA$ and $\PGS$ can be written as
\[
\pi(x) = N(x \mid \bar{m}, \bar{Q}^{-1})\prod_{i = 1}^dg_i(x_i), \quad x \in \R^d,
\]
for some appropriate $x$, $d$, $\bar{m}$ and $\bar{Q}$, 
where $-\log(g_i(x_i))$ is a convex function for every $i$. Then, the results in \cite{ascolani2024entropy} (in particular Theorem $3.1$ therein) imply that the mixing times of a random scan Gibbs sampler on $\pi$ can be upper bounded by the ones of a random scan Gibbs sampler on $N(x \mid \bar{m}, \bar{Q}^{-1})$. This allows to derive explicit expressions, since Gibbs samplers on Gaussian targets are amenable to analytical study (see e.g.\ \citet[Lemma 3.10]{ascolani2024entropy}, and earlier work in \cite{amit1996convergence, roberts1997updating}).

In our context, this means that the mixing time of $\PDA$ is upper bounded by the one of the two-block Gibbs Sampler targeting the corresponding prior distribution, i.e.\  $N(\beta \mid m,Q_0^{-1})N(z \mid X\beta,I_n)$, and similarly for $\PGS$ and the corresponding marginal prior on $z$.
In other words, we can ignore the likelihood because in this context it can only speed up the convergence of both $\PDA$ and $\PGS$.

The results of \cite{ascolani2024entropy} are limited to the random scan case: we apply them directly to $\PGS$ in Section \ref{sec:proof_main} of the Supplementary Material. Extending the approach to the case of two-block deterministic-scan samplers (also called Data Augmentation samplers, such as $\PDA$), requires some technical work, which we carry out in Section \ref{sec:theory}.
Before doing that, we discuss the related literature (Section \ref{sec:literature}) and analyze the implications of Theorem \ref{theo:mixing_times} in various regimes of statistical interest, namely $g$ priors (Section \ref{sec:implications}), random design models (Section \ref{sec:implications}), models with and without the intercept (Section \ref{sec:intercept}); discuss the resulting computational complexity and compare it with the one of gradient-based samplers (Section \ref{sec:comp_cost})
and provide numerical illustrations (Section \ref{sec:simulation}). Finally, we provide some guarantees on using the prior as a starting distribution in Section \ref{sec:starting_distribution}. The code to reproduce the numerical experiments is available at \url{https://github.com/gzanella/ProbitDA}.

\subsection{Related literature}\label{sec:literature}
Other papers have studied the convergence properties of $\PDA$. For example, \cite{roy2007convergence} proved that $\PDA$ is geometrically ergodic through drift and minorization techniques \citep{rosenthal1995minorization}. However, such bounds deteriorate with $n$ and $p$, making them not informative in high-dimensional problems. Subsequent works, and in particular \cite{qin2019convergence, qin2022wasserstein}, showed that such bounds could be substantially improved 
under various assumptions, such as $p$ fixed and $n\to \infty$ with proper assumptions on the data-generating mechanism (Theorem $17$ in \cite{qin2019convergence}), 
$n$ fixed and $p \to \infty$ (Theorem $22$ in \cite{qin2019convergence}), 
as well as other settings such as repeated rows in the  matrix $X$ (see in particular \cite{qin2022wasserstein}).

The most complete bounds up to now are given in the recent work \cite{lee2024fast}, where the authors study the mixing times in Total Variation distance through the profile conductance of $\PDA$. By Pinsker inequality, the $\KL$ bound in \eqref{eq:bound_mix_time_GS} also implies a corresponding bound in Total Variation: in this context the results in Theorems $3.2$ and $3.6$ in \cite{lee2024fast} are similar to ours in terms of overall resulting complexity (as a function of $n$ and $p$), but we employ an arguably easier proof technique (based on \cite{ascolani2024entropy}) that leads to more explicit and smaller constants. Moreover Theorem \ref{theo:mixing_times} depends on the starting distribution $\mu$ through $\KL(\mu, \pi)$, without requiring the stronger assumption that $\mu$ is warm, i.e.\ $\sup_x \, \frac{\d \mu}{\d \pi}(x) < \infty$. In addition, the $\KL$ mixing times bounds in Theorem \ref{theo:mixing_times} follow by the stronger result we prove on the one-step entropy contraction: see Section \ref{sec:theory} for details. This implies a lower bound on the spectral gap (see \cite{caputo2023lecture} below Lemma $2.15$) and upper bounds on $\chi^2$-mixing times (see Remark \ref{rmk:results}). 

On the contrary, we found no explicit results on the convergence of $\PGS$. Thus the bound in \eqref{eq:bound_mix_time_PCG}, again based on \cite{ascolani2024entropy} (see Section \ref{sec:proof_main} of the Supplementary Material), is novel to the best of our knowledge.

\subsection{Implications}\label{sec:implications}

We now consider two popular choices of $Q_0$ and investigate the implications of Theorem \ref{theo:mixing_times}.

\subsubsection{g prior} If $X^TX$ is invertible, then the so-called $g$ prior \citep{zellner1986assessing, liang2008mixtures} is given by $Q_0^{-1} = g(X^TX)^{-1}$, with $g\in(0,\infty)$ being a multiplicative scalar. 
The $g$ prior requires $X^TX$ to be invertible, which for example cannot hold when $p>n$. 
We thus consider the more general case $Q_0^{-1} = (X^TX/g + cI_p)^{-1}$ with $c \geq 0$, which is always well defined if $c>0$ and reduces to the standard g prior if $c=0$. 
In the next corollary we obtain upper bounds on the mixing times for those cases.
\begin{corollary}\label{crl:gpriors}
Let $Q_0^{-1} = (X^TX/g + cI_p)^{-1}$, with $c\geq 0$. Then
\[
\tmix(\epsilon,\mu, \PDA) \leq (2+g)\log\left(\KL(\mu,\pi)/\epsilon\right), \quad \tmix(\epsilon,\mu_1, \PGS^n) \leq \left(1+g\right)\log\left(\KL(\mu,\pi)/\epsilon\right)\,.
\]
\end{corollary}
\begin{proof}
We have $XQ_0^{-1}X^T = X(X^TX/g + cI_p)^{-1}X^T=gX(cgI_p+X^TX)^{-1}X^T$.
By Woodbury's identity
$$
X(cgI_p+X^TX)^{-1}X^T
=I_n-(I_n+XX^T/(cg))^{-1}\,.
$$ 
The above equalities imply $\lambda_{max}(XQ_0^{-1}X^T) \leq g$ and the bounds follow from Theorem \ref{theo:mixing_times}.
\end{proof}
Interestingly the upper bounds in Corollary \ref{crl:gpriors} do not depend on $X$, nor on $n$ and $p$. This implies that convergence speed does not deteriorate in high dimensions if $g$ is held fixed. On the other hand, the bounds increase with $g$, i.e.\ as the prior becomes less informative. 
These features are confirmed by the simulations, and they occur not only for worst-case data $y$ but also under randomly generated $y$, see Section \ref{sec:simulation}.

\subsubsection{Diagonal precision under random design}

We now consider the case of isotropic prior and random design matrix $X=(X_{ij})_{ij}\in\R^{n\times p}$, as specified in the next assumption.
\begin{asmp}
\label{asmp:no_int}
Assume either:\\
(a) $Q_0^{-1}=cI_p$ and $X_{ij} = G_{ij}/\sqrt{p}$ or \\
(b) $Q_0^{-1}=(c/p)I_p$ and $X_{ij}=G_{ij}$,\\
with $c>0$ and $G_{ij} \simiid F$ for $(i,j)\in\{1,\dots,n\}\times\{1,\dots,p\}$, where $F \in \sP(\R)$ has zero mean, unit variance and finite fourth moment.
\end{asmp}
Rescaling either $Q_0^{-1}$ or $X_{ij}^2$ by $1/p$, as in Assumption \ref{asmp:no_int},  is a standard practice in high-dimensional Bayesian regression \citep{simpson2017penalising, fuglstad2020intuitive}, which ensures that the variance of linear predictors $x_i^T\beta$ under the prior remains roughly constant as $p$ increases, since $\text{Var}(x_i^T\beta) =
\frac{c}{p}\sum_{j = 1}^pG_{ij}^2\to c$ almost surely as $p\to\infty$. 

The random design assumption allows us to use random matrix theory to obtain high-probability bounds on $\lambda_{max}(XX^T)$ and $\lambda_{min}(XX^T)$, and thus on mixing times, as detailed in the next corollary.
\begin{corollary}\label{crl:noint}
Under Assumption \ref{asmp:no_int} we have that
\[
\tmix(\epsilon,\mu, \PDA) \leq \left[ 2+c(1+\sqrt{r})^2+\delta\right]\log\left(\KL(\mu,\pi)/\epsilon\right)
\]
and
\[
\tmix(\epsilon,\mu_1, \PGS^n) \leq \frac{1+ c(1+\sqrt{r})^2+\delta}{1+c(1-\sqrt{\min\{1,r\}})^2}\log\left(\KL(\mu,\pi)/\epsilon\right),
\]
almost surely as $n \to \infty$ and $n/p \to r \in (0, \infty)$, for any arbitrarily small positive constant $\delta > 0$.
\end{corollary}
\begin{proof}
Denoting $G=(G_{ij})_{ij}\in\R^{n\times p}$, Theorem $2$ in \cite{bai2008limit} implies that
\begin{equation}\label{eq:bound_eigenvalues}
\lambda_{max}(XQ_0^{-1}X^T) =\frac{c}{p}\lambda_{max}(GG^T) \to c(1+\sqrt{r})^2,
\end{equation}
almost surely as $n \to \infty$ and $n/p \to r \in (0, \infty)$, and, if $r < 1$, also $\lambda_{min}(XQ_0^{-1}X^T) \to c(1-\sqrt{r})^2$ almost surely. Combining  those with Theorem \ref{theo:mixing_times} gives the result.
\end{proof}
The results of Corollary \ref{crl:noint} provide various insights, namely:
\begin{enumerate}
    \item[(i)] Both $\PDA$ and $\PGS$ mix fast (i.e.\ in $\mathcal{O}(1)$ iterations) in high-dimensional scenarios where $p$ is comparable to (or larger than) $n$ and $c$ is small. 
    \item[(ii)] When $p < n$ the bound on $\PGS$ is similar to the one on $\PDA$.
When $p > n$ and $c$ is large, the bound on $\PGS$ can be better. 
This dependence on $c$ and $r$ is also confirmed by numerical experiments, see Section \ref{sec:comparison_supp} in the Supplementary Material.

\item[(iii)] If $c$ is fixed, both bounds deteriorate when $n/p$ grows. 
\end{enumerate}

These features are empirically confirmed by the simulation study in Section \ref{sec:simulation}.

\section{The role of the intercept (and unbalanced data)}\label{sec:intercept}

Contrarily to Assumption \ref{asmp:no_int}, where each column is rescaled by a factor of $1/\sqrt{p}$, we now assume that the first column of $X$ has all entries equal to $1$, i.e.\ that $\beta_1$ is the intercept. It is indeed common practice not to rescale the intercept (see e.g.\ \cite[Section 2]{sardy2008practice}), or equivalently not to strongly shrink it towards $0$,
in order to allow the intercept to account for the average proportion of ones in $y$ marginally over the covariates. 
We state the assumption only in terms of rescaling covariates for the sake of brevity.
\begin{asmp}
\label{asmp:with_int}
$Q_0^{-1}=cI_p$, 
$X_{i1}=1$ for $i\in\{1,\dots,n\}$, $X_{ij} = G_{ij}/\sqrt{p}$ and $G_{ij} \simiid F$ 
for $i\in\{1,\dots,n\}$ and $j\in\{2,\dots,p\}$, where $c>0$ and $F \in \sP(\R)$ has zero mean, unit variance and finite fourth moment.
\end{asmp}
Including the intercept has a major impact on the mixing of $\PDA$ and $\PGS$, 
as shown below.
\begin{corollary}\label{crl:bound_with_int}
Under Assumption \ref{asmp:with_int} we have that
\[
\tmix(\epsilon,\mu, \PDA) \leq (2+(c+\delta) n)\log\left(\KL(\mu,\pi)/\epsilon\right)
\]
and
\[
\tmix(\epsilon,\mu_1,\PGS^n) \leq (1+(c+\delta) n)\log\left(\KL(\mu,\pi)/\epsilon\right),
\]
almost surely as $n \to \infty$ and $n/p \to r \in (0, \infty)$, for any arbitrarily small positive constant $\delta > 0$.
\end{corollary}
\begin{proof}
By construction the matrix $X^TX$, whose non-zero eigenvalues are the same of $XX^T$, has the form
\[
X^TX =
\begin{bmatrix}
    n & \sum_{i = 1}^nX_{i2} & \dots &\sum_{i = 1}^nX_{ip}\\
    \sum_{i = 1}^nX_{i2}& 0 & \dots & 0\\
    \vdots & &\\
    \sum_{i = 1}^nX_{ip}& 0 & \dots & 0
\end{bmatrix} +
\begin{bmatrix}
    0 & 0 & \dots &0\\
    0& \\
    \vdots & &\tilde{X}^T\tilde{X}\\
    0
\end{bmatrix}
\]
where $\tilde{X} \in \R^{n\times p-1}$ is the matrix $X$ without the first column. Then by Weyl's inequality, we have that
$\lambda_{max}(X^TX) \leq n + \lambda_{max}(\tilde{X}^T\tilde{X})$ and $\lambda_{max}(\tilde{X}^T\tilde{X}) \to (1+\sqrt{r})^2$ almost surely by Theorem $2$ in \cite{bai2008limit}. Thus $\lambda_{max}(XQ_0^{-1}X^T) \leq (c+\delta)n$ almost surely as $n \to \infty$ and $n/p\to r$ for every $\delta>0$, and the result follows by Theorem \ref{theo:mixing_times}.
\end{proof}
The bounds in Corollary \ref{crl:bound_with_int} deteriorate linearly with $n$, implying that $\sO(n)$ iterations are sufficient for convergence. The next proposition, in the intercept-only case (i.e.\ $p = 1$), shows that this is not improvable in general and it corresponds to the case of highly imbalanced data. The proof can be found in Section \ref{sec:proof_lower_bound} of the Supplementary Material.
\begin{proposition}\label{prop:lower_bound_KL}
Let $p = 1$, $m = 0$, $Q_0^{-1} = c>0$, $x_i = 1$ for every $i\in\{1,\dots,n\}$ and $y_i = 1$ for every $i$ (or $y_i = 0$ for every $i$). Then, for every $n\geq 8/c$ 
\begin{enumerate}
\item[(i)] we have that
\[
\sup_{\mu} \,\frac{\KL(\mu \PDA, \pi)}{\KL(\mu, \pi)} \geq 1- \frac{\log(cn)}{d(1+cn)}
\]
for a universal constant $d >0$, where the supremum is taken over every $\mu \in \sP(\R^n \times \R)$ such that $\KL(\mu, \pi) < \infty$.

\item[(ii)] there exists $\mu \in \sP(\R^n \times \R)$ such that
\begin{align}\label{eq:lower_t_mix_two}
(3 + 2cn) \log\left( \frac{\chi^2( \mu, \pi)}{\epsilon}\right)   \geq \tau_{\mathrm{mix},2}(\epsilon, \mu, \PDA)
    \geq 
    d'\left(\frac{1+cn}{\log (cn)}\right)\log\left( \frac{\chi^2( \mu, \pi)}{\epsilon}\right)
\end{align}
for a universal constant $d' >0$  and all $\epsilon>0$.
\end{enumerate}
\end{proposition}
\begin{remark}
Proposition \ref{prop:lower_bound_KL} is inspired by results in \cite{johndrow2019mcmc}, which prove that $\PDA$ with imbalanced data and intercept only mixes slowly as $n$ increases. In particular Theorem $3.2$ therein implies a lower bound of order $\sqrt{n}$ on the mixing times in Total variation distance. Here we improve this to order $n$ (ignoring logarithmic terms) by passing to the $\chi^2$ divergence.
\end{remark}
The proof of Proposition \ref{prop:lower_bound_KL} relies on showing that Var$(\beta_1 \mid z) = \sO(n^{-1})$  for every $z$ while  $\text{Var}_\pi(\beta_1)=\sO(1)$ in the imbalanced case as $n\to\infty$. Part $(i)$ then shows that in the worst case $\PDA$ reduces the entropy by a factor of $1/n$ in one iteration, suggesting that $\sO(n)$ iterations may be needed to reach stationarity; this is confirmed in part $(ii)$, which provides matching lower and upper bounds on mixing times in $\chi^2$ divergence.

In order to solve this issue, we consider a simple modification of Algorithm \ref{alg:PDA}, where an additional Metropolis update of $\beta_1$ from $\pi(\beta_1 \mid \beta_{-1})$ is included before updating $z$: see Algorithm \ref{alg:PDA2} for the pseudocode. 
The resulting algorithm is still $\pi(z,\beta)$-invariant (as, for example, it can be interpreted as an instance of partially-collapsed Gibbs sampler \citep{van2008partially}).
Note that the additional update of $\beta_1$ is invariant with respect to $\pi(\beta_1 \mid \beta_{-1})$, which we expect to have $\sO(1)$ variance in the imbalanced case, instead of $\pi(\beta_1\mid z)$, which has always $\sO(n^{-1})$ variance. This modification (which recalls the one proposed in \cite{johndrow2019mcmc} for $p = 1$) is shown empirically to mix fast in both balanced and imbalanced settings in Section \ref{sec:simulation}.
Alternative solutions have been explored in the literature, such as interweaving strategies \citep{yu2011center} and parameter expansions \citep{zens2024ultimate}, which we also expect to be effective in solving the same issue.

\begin{algorithm}[htbp]
\begin{algorithmic}
\State Initialize $(z^{(0)},\beta^{(0)})$. 
\For{$t \geq 1$}
\State Set $\tilde{\beta} = \beta^{(t-1)}$.
\State Sample $\beta_1 \sim N(\tilde{\beta}_1, \sigma^2)$ and set $\tilde{\beta} = (\beta_1,\tilde{\beta}_{-1})$ w.p. $\min\left\{1, \pi(\beta_1,\tilde{\beta}_{-1})/\pi(\tilde{\beta}) \right\}$.

\State Sample $z^{(t)}_i \sim \pi(z_i \mid \tilde{\beta})\, \propto \, N(z_i \mid x_i^T\tilde{\beta}, 1)\mathbbm{1}(y_i = g(z_i))$, for $i = 1, \dots, n$.

    \State Sample $\beta^{(t)} \sim \pi(\beta \mid z^{(t)})$, with $\pi(\beta \mid z)$ as in \eqref{eq:posterior_beta}.

   \EndFor
\end{algorithmic}
\caption{(Modified data augmentation Gibbs sampler $\PDAmod$)
\label{alg:PDA2}}
\end{algorithm}

Let us stress again that the results in Theorem \ref{theo:mixing_times} are worst-case with respect to $y$, and may substantially differ from the average case: for example, we expect that if the dataset is balanced (i.e.\ if $n^{-1}\sum_{i = 1}^ny_i$ is far from $0$ and $1$) then both $\PDA$ and $\PGS$ converge fast in the setting of Proposition \ref{prop:lower_bound_KL}: indeed also $\text{Var}_\pi(\beta_1)=\sO(n^{-1})$ is expected in that case. This is coherent with the findings in \cite{qin2019convergence}, which show that if $p$ is fixed and $n \to \infty$ with data generated according to model \eqref{eq:probit_DA} the mixing times remain bounded with $n$, and with the simulations in Section \ref{sec:simulation}.

\section{Computational cost and comparisons}\label{sec:comp_cost}
We now complement the above mixing time bounds with a discussion on the overall computational cost of $\PDA$ and $\PGS$, and a brief comparison with gradient-based samplers. 

\subsection{Cost per iteration}

Running either $\PDA$ or $\PGS^{n}$ requires 
$\sO(np\min\{n,p\})$ pre-computation cost\footnote{Note that the pre-computation cost refers to a \emph{single} matrix factorization or inversion which, while being nominally of order $\sO(np\min\{n,p\})$, is usually not the dominant cost in practice.  
Moreover, in settings where this becomes the dominant cost, there is a large body of well-established tools that could be used to reduce this cost at the price of some small approximation error, see e.g.\ \cite{pandolfi2024conjugate} for examples of using conjugate gradient solvers to avoid full matrix inversions in Gibbs Samplers with Gaussian conditionals. To make these arguments complete, one should then quantify how the approximation error transfer into the posterior one, which we leave to future work.} to compute and factorize the covariance matrix of $\beta$, which  needs to be done once, and $\sO(n p)$ cost per iteration. 
When $p>n$ the cost per iteration can be reduced to $\sO(n^2)$ in some cases, see Section $A$ of the Supplementary Material for more details on this and the actual implementation.

\subsection{Comparison with gradient-based schemes} An alternative to $\PDA$ or $\PGS$ is to target directly $\pi(\beta)$ in \eqref{eq:model} with a gradient-based MCMC algorithm, such as Langevin or Hamiltonian Monte Carlo, without relying on the data augmentation structure in \eqref{eq:posterior}. Indeed $\pi(\beta)$ is log-concave and a large literature is available on the resulting performances of gradient-based samplers in this setting  (see e.g.\  \cite{chewi2023log} for an overview). 
Computing the gradient of $\log\pi(\beta)$ requires $\sO(np)$ cost, which matches the cost per iteration of $\PDA$ and $\PGS^{n}$. To the best of our knowledge, currently available upper bounds on the mixing times of gradient-based MCMC schemes which apply to $\pi(\beta)$ are of the form $\sO(\kappa^a p^b)$, where $\kappa$ is the condition number of $\pi(\beta)$ and both $a \geq 1$ and $b > 0$ depend on the specific algorithm. 
For example, \cite{wu2022minimax, altschuler2024faster} proved that the mixing time of the Metropolis-Adjusted Langevin Algorithm (MALA), possibly after an algorithmic warm start, is of order $\sO(\kappa p^{1/2})$. 
Proposition \ref{prop:cond_number} in the Supplementary Material shows that,
after pre-conditioning with the prior variance $Q_0^{-1}$, the condition number of $\pi(\beta)$ satisfies $\kappa \leq 1 + \lambda_{max}(XQ_0^{-1}X^T)$. This implies an upper bound of order $\sO\big( p^{1/2} \big(1 + \lambda_{max}(XQ_0^{-1}X^T)\big)\big)$ for the mixing times for MALA, which is strictly worse than the upper bounds for $\PDA$ and $\PGS$ in Theorem \ref{theo:mixing_times} by a factor of $p^{1/2}$. The latter can be interpreted as the additional cost due to the discretization of the Langevin diffusion, see e.g.\ \citet[Section 4.2]{ascolani2024entropy} for more discussion on this.

\section{Simulations}\label{sec:simulation}

\subsection{Practical considerations: coupling-based upper bounds}

In order to empirically assess the above bounds, we rely on recent couplings methodologies \citep{jacob2020unbiased, biswas2019estimating}, which allow to generate numerical upper bounds to the total variation (TV) distance $\lTV \mu P^t -\pi \rTV = \sup_A |\mu P^t(A) -\pi(A)|$. 
In particular, we consider the methodology introduced in \cite{biswas2019estimating}, which we briefly describe. Consider a bivariate Markov chain $(X^{(t)}_1, X^{(t)}_2)_t$ with operator $K((x_1, x_2), \cdot)$ such that marginally $(X^{(t)}_i)_t$ is a Markov chain with kernel $P$ for $i = 1,2$. The kernel $K$ is called a coupling and it is usually chosen so that the meeting time $\tau^{(L)} = \inf\{t > L \mid X^{(t)}_1 =X^{(t-L)}_{2}  \}$ is almost surely finite and $X^{(t)}_1 = X^{(t-L)}_{2}$ for every $t > \tau^{(L)}$, where $L > 0$ is a suitable lag. The pseudocode to sample a realization of $\tau^{(L)}$ (which corresponds to Algorithm $1$ in \cite{biswas2019estimating}), is given in Algorithm \ref{alg:meeting_time}.

\begin{algorithm}[htbp]
\begin{algorithmic}
\State Initialize $X^{(0)}_2 \sim \mu$, $X^{(0)}_1 \sim \mu$ and $X^{(t)}_1\mid X^{(t-1)}_{1} \sim P(X^{(t-1)}_{1}, \cdot)$ for $t = 1, \dots, L$. 
\For{$t > L$}
    \State Sample $\left(X^{(t)}_1, X^{(t-L)}_{2} \right) \sim K\left((X^{(t-1)}_{1},X^{(t-L-1)}_{2}), \cdot\right)$
    \State If $X^{(t)}_1 = X^{(t-L)}_{2}$, return $\tau^{(L)} = t$ and exit the for loop.
   \EndFor
\end{algorithmic}
\caption{(Sampling meeting times $\tau^{(L)}$)
\label{alg:meeting_time}}
\end{algorithm}
Theorem $1$ in \cite{biswas2019estimating} shows that $\lTV \mu P^t -\pi \rTV \leq \bar d (t)$ with
\begin{equation}\label{eq:bound_TV}
\bar d (t)=\mathbb{E}\left[\max\left\{0, \left\lceil \frac{\tau^{(L)}-L-t}{L} \right\rceil \right\} \right],
\end{equation}
for every $t$, where the bound is tighter as $L$ increases. Thus, we can use Algorithm \ref{alg:meeting_time} to obtain $N$ independent realizations of $\tau^{(L)}$ and approximate the right hand side of \eqref{eq:bound_TV} with their empirical average. As regards the choice of $K$, we consider the two-step coupling described in \cite{ceriani2024linear} (see also references therein): when the two chains are far (i.e.\ $d(X^{(1)}_t, X^{(2)}_{t-L}) > \epsilon$, for some suitable distance function $d$) then a contractive coupling is used in order to make the chains closer, otherwise a maximal coupling (maximizing the probability of the chains being exactly equal in one step) is employed. Section $C$ in the Supplementary Material provides more details and a pseudocode for the couplings we employ.
The results is a Monte Carlo estimate of $\bar d (t)$, which we can either plot as a function of $t$ to monitor convergence (as in Figure \ref{fig:withint1}) or use to upper bound the TV-mixing times, $\tmix^{TV}(\epsilon, \mu,P) 
=\inf\{t\geq 1\,:\,\lTV \mu P^t -\pi \rTV \leq \epsilon\}$, as
\begin{equation}\label{eq:bound_TV_mix}
\tmix^{TV}(\epsilon, \mu,P)
\leq
\inf\left\{t\geq 1\,:\,
 \bar{d}(t)\leq \epsilon\right\}\,,
\end{equation}
which follows from $\lTV \mu P^t -\pi \rTV \leq \bar d (t)$.

Recall that, while our results in Theorem \ref{theo:mixing_times} bound mixing times in KL, by the Pinsker inequality they also provide bounds to mixing times in TV that are equivalent up to small multiplicative constants (see e.g.\ Remark 3.6 in \cite{ascolani2024entropy}). In this sense, looking at TV-mixing times can also be seen as a way to validate the tightness of the bounds in Theorem \ref{theo:mixing_times}.

\subsection{Simulation studies for various priors}
In Tables \ref{table:imbalanced} and \ref{table:random} we report the upper bounds to $\tmix^{TV}(\epsilon, \mu,P)$ based on \eqref{eq:bound_TV_mix}, with $\mu$ as in Section \ref{sec:starting_distribution} below. 
The rows refer to different choices of the design matrix $X$ and prior precision $Q_0$, while different columns feature distinct combinations of $n$ and $p$. Table \ref{table:imbalanced} considers the imbalanced case with $y_i = 1$, while for Table \ref{table:random} the responses are generated from the model itself, as defined in \eqref{eq:model_orig}.

\begin{table}[]
\centering
\setlength{\tabcolsep}{2pt}
\renewcommand{\arraystretch}{0.9}
\begin{tabular}{c|c|ccc|ccc|ccc}
\hline
\multicolumn{1}{c|}{\multirow{2}{*}{\begin{tabular}[c]{@{}c@{}}Imbalanced data:\\$y_i=1$ for all $i$\end{tabular}}} &
\multicolumn{1}{c|}{\multirow{2}{*}{Method}} &
\multicolumn{3}{c|}{n/p=0.2} &
\multicolumn{3}{c|}{n/p=1.25} &
\multicolumn{3}{c}{n/p=3} \\ \cline{3-11} 
\multicolumn{1}{c|}{} &
\multicolumn{1}{c|}{} &
\multicolumn{1}{c}{p=50} &
\multicolumn{1}{c}{p=100} &
\multicolumn{1}{c|}{p=200} &
\multicolumn{1}{c}{p=50} &
\multicolumn{1}{c}{p=100} &
\multicolumn{1}{c|}{p=200} &
\multicolumn{1}{c}{p=50} &
\multicolumn{1}{c}{p=100} &
\multicolumn{1}{c}{p=200} \\ \hline
\multirow{2}{*}{\begin{tabular}[c]{@{}c@{}}g prior\\ $(g=1, c = 0.001)$\end{tabular}} &
$\PDA$ & 11
& 11
& 11
& 6
& 7
& 7
& 6
& 6
& 6
\\
 & $\PGS^n$ & 8
& 10 
& 11
& 20
& 23
& 24
& 24
& 25
& 27
\\\hline
\multirow{2}{*}{\begin{tabular}[c]{@{}c@{}}g prior\\ $(g=10, c = 0.001)$\end{tabular}} &
$\PDA$ & 57
& 62
& 65
& 38
& 40
& 43
& 27
& 29
& 29
\\
& $\PGS^n$ & 11
& 13
& 15
& 34
& 36
& 38
& 46
& 48
& 49
\\\hline
\multirow{2}{*}{\begin{tabular}[c]{@{}c@{}}Assumption \ref{asmp:no_int}\\ ($c=1$)\end{tabular}} &
$\PDA$ & 6
& 7
& 7
& 7
& 8
& 8
& 9
& 9
& 9
\\
&$\PGS^n$ & 14
& 16
& 18
& 22
& 24
& 26
& 26
& 27
& 30
\\\hline
\multirow{2}{*}{\begin{tabular}[c]{@{}c@{}} Assumption \ref{asmp:no_int}\\ ($c=10$)\end{tabular}} &
$\PDA$ & 38
& 43
& 44
& 39
& 44
& 54
& 33
& 24
& 20
\\
&$\PGS^n$ & 16
& 19
& 22
& 36
& 38
& 45
& 52
& 42
& 39
\\\hline
\multirow{2}{*}{\begin{tabular}[c]{@{}c@{}}Assumption \ref{asmp:with_int}\\ ($c=1$)\end{tabular}} &
$\PDA$ & 21
& 35
& 56
& 81
& 143
& 247
& 160
& 302
& 591
\\
 &
$\PGS^n$ 
& 26
& 40
& 62
& 102
& 169
& 301
& 244
& 416
& 724
\end{tabular}
\caption{Upper bounds on the mixing times $\tmix^{TV}(\epsilon, \mu,\PDA)$ and $\tmix^{TV}(\epsilon, \mu,\PGS^n)$, for $\epsilon=0.1$ and $\mu(\d z, \d \beta) = N(\d\beta \mid 0, Q_0^{-1})\pi(\d z \mid \beta)$, obtained from \eqref{eq:bound_TV_mix}, taking $L=200$ and estimating $\bar{d}(t)$ with $N = 500$ independent simulations of $\tau^{(L)}$. $X$ is generated either as in Assumption \ref{asmp:with_int} (rows 1, 2 and 5) or as in Assumption \ref{asmp:no_int} (rows 3 and 4), with $F= N(0,1)$. In all cases, $y_i = 1$ for $i\in\{1,\dots,n\}$.
}\label{table:imbalanced}
\end{table}

\begin{table}[]
\centering
\setlength{\tabcolsep}{2pt}
\renewcommand{\arraystretch}{0.9}
\begin{tabular}{c|c|ccc|ccc|ccc}
\hline
\multicolumn{1}{c|}{\multirow{2}{*}{\begin{tabular}[c]{@{}c@{}}Well-specified data:\\$y_i\sim \text{Bern}(\Phi(x_i^T\beta))$\end{tabular}}} &
\multicolumn{1}{c|}{\multirow{2}{*}{Method}} &
\multicolumn{3}{c|}{n/p=0.2} &
\multicolumn{3}{c|}{n/p=1.25} &
\multicolumn{3}{c}{n/p=3} \\ \cline{3-11} 
\multicolumn{1}{c|}{} &
\multicolumn{1}{c|}{} &
\multicolumn{1}{c}{p=50} &
\multicolumn{1}{c}{p=100} &
\multicolumn{1}{c|}{p=200} &
\multicolumn{1}{c}{p=50} &
\multicolumn{1}{c}{p=100} &
\multicolumn{1}{c|}{p=200} &
\multicolumn{1}{c}{p=50} &
\multicolumn{1}{c}{p=100} &
\multicolumn{1}{c}{p=200} \\ \hline
\multirow{2}{*}{\begin{tabular}[c]{@{}c@{}}g prior\\ $(g=1, c = 0.001)$\end{tabular}} &
$\PDA$ & 11
& 11
& 11
& 6
& 7
& 7
& 5
& 6
& 6
\\
 & $\PGS$ & 8
& 10
& 11
& 20
& 22
& 24
& 23
& 24
& 26
\\\hline
\multirow{2}{*}{\begin{tabular}[c]{@{}c@{}}g prior\\ $(g=10, c = 0.001)$\end{tabular}} &
$\PDA$ & 58
& 63
& 64
& 40
& 41
& 44
& 25
& 28
& 30
\\
& $\PGS$ & 11
& 13
& 15
& 34
& 36
& 37
& 43
& 46
& 49
\\\hline
\multirow{2}{*}{\begin{tabular}[c]{@{}c@{}}Assumption \ref{asmp:no_int}\\ ($c=1$)\end{tabular}} &
$\PDA$ & 6
& 7
& 7
& 8
& 9
& 9
& 11
& 11
& 11
\\
&$\PGS$ & 13
& 16
& 18
& 22
& 25
& 26
& 28
& 29
& 31
\\\hline
\multirow{2}{*}{\begin{tabular}[c]{@{}c@{}}Assumption \ref{asmp:no_int}\\ ($c=10$)\end{tabular}} &
$\PDA$ & 38
& 43
& 47
& 58
& 71
& 69
& 106
& 104
& 90
\\
 &
$\PGS$ 
& 16
& 19
& 21
& 48
& 55
& 54
& 130
& 133
& 118
\\\hline
\multirow{2}{*}{\begin{tabular}[c]{@{}c@{}}Assumption \ref{asmp:with_int}\\ ($c=1$)\end{tabular}} &
$\PDA$ & 21
& 9
& 10
& 30
& 10
& 34
& 27
& 13
& 17
\\
 &
$\PGS^n$  & 26
& 18
& 20
& 45
& 25
& 52
& 49
& 31
& 36
\end{tabular}
\caption{Same as Table \ref{table:imbalanced} with data generated from the correct model, i.e.\ $y_i\sim \text{Bern}(\Phi(x_i^T\beta))$ for all $i\in\{1,\dots,n\}$, with $\beta \sim N(0, Q_0^{-1})$.}\label{table:random}
\end{table}

The first two rows refer to $g$ priors, with different values of the parameter $g$ which measures the amount of prior information. Coherently with Corollary \ref{crl:gpriors}, the estimates of the mixing times are always small, regardless of $n$, $p$ and the data generation mechanism: interestingly, there seems to be very little variation for different choices of $n$ and $p$ with the same ratio $n/p$. Moreover, as expected, the mixing times increase with $g$ (i.e.\ passing from the first to the second row) in all the scenarios.

Third and fourth rows consider isotropic priors with random $X$ and no intercept, i.e.\ Assumption \ref{asmp:no_int}. Also here the estimates are quite small, with an increasing trend in $r$ as suggested by Corollary \ref{crl:noint}: the latter phenomenon is more apparent with data generated from the model and $c$ is large. Also, the mixing times increases with $c$, going from the third to the fourth row. Such increase is negligible for $\PGS$ when $r < 1$, which is coherent with the bound in Corollary \ref{crl:noint}.

\begin{figure}
\centering
\includegraphics[width=.48\textwidth]{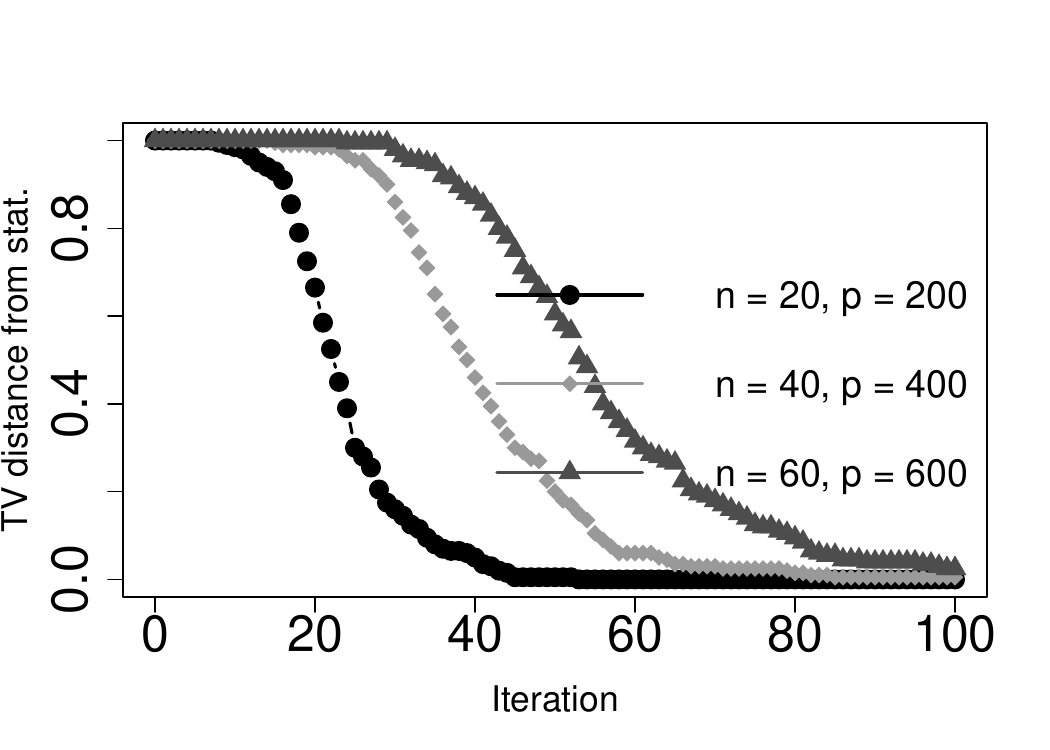} \quad
\includegraphics[width=.48\textwidth]{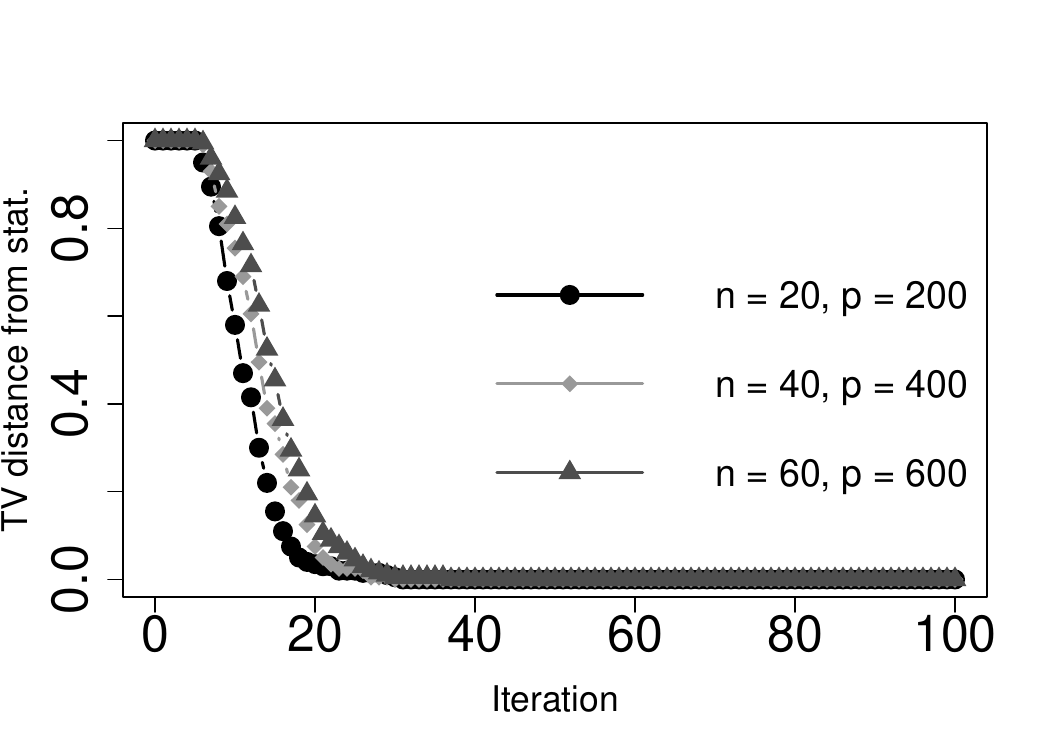}\\
\includegraphics[width=.48\textwidth]{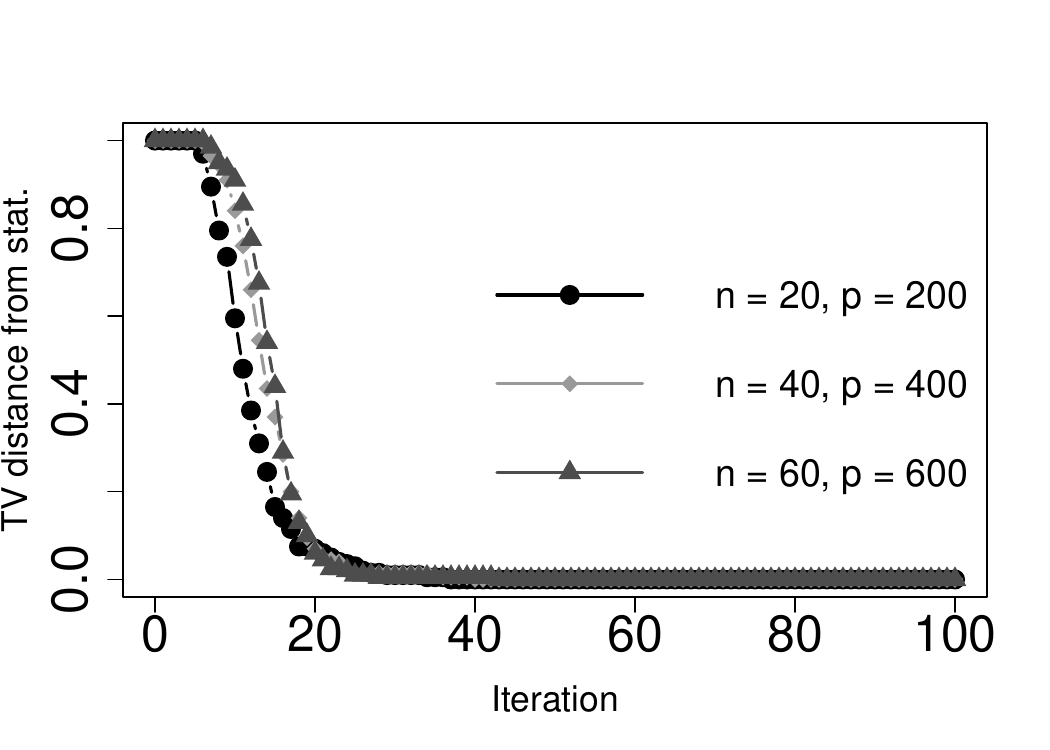} \quad
\includegraphics[width=.48\textwidth]{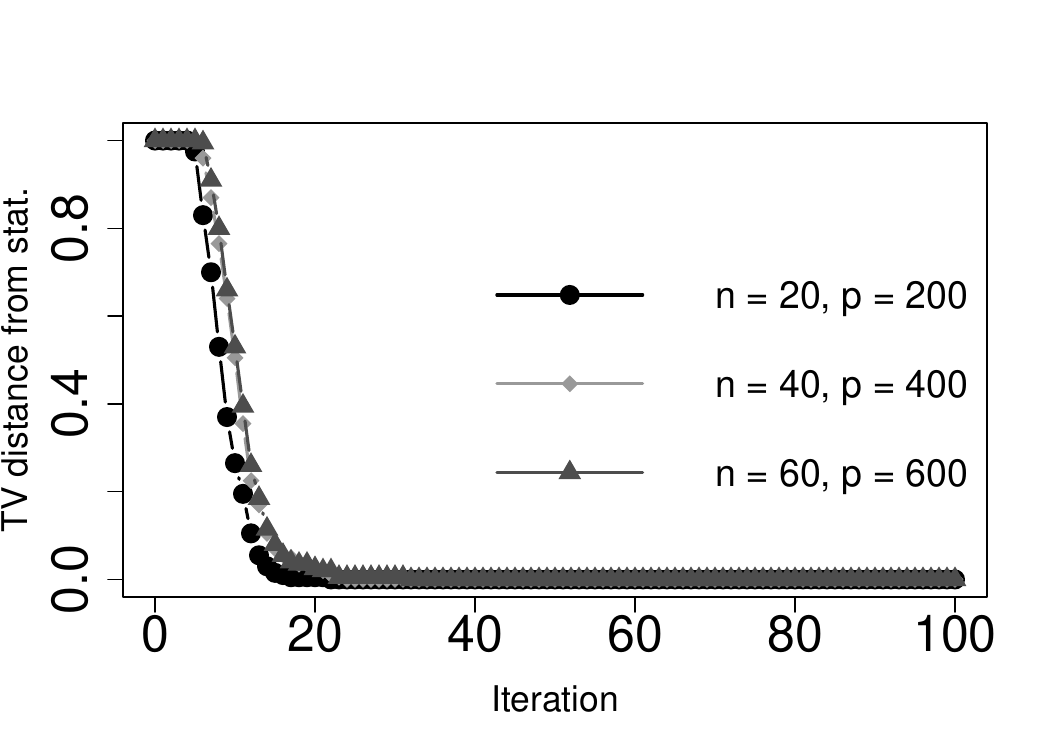}
 \caption{\small{
Upper bounds on 
$\lTV \mu \PDA^t -\pi \rTV$  (left column) and $\lTV \mu \PDAmod^t -\pi \rTV$ (right column) as a function of $t$, with $\PDAmod$ defined in Algorithm \ref{alg:PDA2}, $\mu(\d z, \d \beta) = N(\d\beta \mid 0, I_p)\pi(\d z \mid \beta)$, $Q_0^{-1} = I_p$ and $X$ generated according to Assumption \ref{asmp:with_int} with $F = N(0,1)$. Bounds are obtained from \eqref{eq:bound_TV}, taking $L=500$ and estimating $\bar{d}(t)$ with $N = 500$ independent simulations of $\tau^{(L)}$. Observations are generated as $y_i = 1$ (top row) and according to model \eqref{eq:probit_DA} (bottom row).
  }}
 \label{fig:withint1}
\end{figure}

The last row considers the case with intercept, as in Assumption \ref{asmp:with_int}. Here the two tables yield very different behaviour: with imbalanced data (Table \ref{table:imbalanced}) the mixing times grow quickly with $n$ regardless of $p$, while for random data (Table \ref{table:random}) they do not. This empirically confirms Corollary \ref{crl:bound_with_int}, which suggests that mixing times increases with $n$ in the worst case. The behaviour for random data was also expected given previous findings \citep{qin2019convergence}. 
To confirm that the problem is given only by the intercept, as suggested by Proposition \ref{prop:lower_bound_KL}, Figure \ref{fig:withint1} reports the Total Variation bounds for $\PDA$ (first column) and $\PDAmod$ (second column) defined in Algorithm \ref{alg:PDA2}. When the data are imbalanced (first row), the former quickly deteriorates with $n$ while the latter remain unaffected. With random data (second row) no notable effect is visible for both algorithms.

\section{Discussion}\label{sec:discussion}

\subsection{Some practical takeaways (and a computationally convenient prior)}\label{sec:take_aways}
The results of this paper provide guidance on which choices of $Q_0$ are expected to lead to fast mixing of $\PDA$ and $\PGS^n$, or lack thereof. 
In particular, Corollary \ref{crl:noint} implies that, under Assumption \ref{asmp:no_int}, choosing $c=b/(1+r)$, with $r=n/p$ and $b>0$ fixed, leads to mixing times that are bounded with respect to both $n$ and $p$, as formalized below.
\begin{corollary}\label{crl:practical}
Under Assumption \ref{asmp:no_int} and $c=b/(1+r)$ with $b>0$, we have
$\tmix(\epsilon,\mu, \PDA) \leq (2+2b)\log\left(\KL(\mu,\pi)/\epsilon\right)
$ and $\tmix(\epsilon,\mu_1, \PGS^n) \leq (1+2b)\log\left(\KL(\mu,\pi)/\epsilon\right)
$ 
almost surely as $n \to \infty$ and $n/p \to r \in (0, \infty)$.
\end{corollary}
\begin{proof}
Follows from Corollary \ref{crl:noint} and
$c(1+\sqrt{r})^2+\delta
=
b(1+\sqrt{r})^2/(1+r)+\delta
<
2b$ for small enough $\delta>0$, which follows from $(1+\sqrt{r})^2< 2(1+r)$ for $r>0$.
\end{proof}

When covariates are standardized to unit variance, the choice $c=b/(1+r)$ corresponds to setting
\begin{align}\label{eq:prior_var_minimal}
Q_0^{-1} = \frac{c}{p}I_p= \frac{b}{p+n}I_p\,.
\end{align}
We can interpret \eqref{eq:prior_var_minimal} as follows:
while rescaling $Q_0^{-1}$ by $p^{-1}$ is natural for statistical reasons (see e.g.\ discussion and references after Assumption \ref{asmp:no_int}), rescaling it also by $n^{-1}$ is computationally convenient since it guarantees fast convergence of $\PDA$ and $\PGS^n$. 
When $r=n/p$ is small, the latter modification \eqref{eq:prior_var_minimal} is equivalent to the standard $p^{-1}$ scaling of $Q_0^{-1}$, while when $r\gg 1$ it increases the amount of shrinkage or informativity of the prior (roughly speaking keeping it as a constant, even if possibly small, fraction of the data informativity). 
Exploring the statistical properties of such a choice is beyond the scope of this work, and we leave it to future research.

Combined with the results of Section \ref{sec:intercept}, \eqref{eq:prior_var_minimal} leads to the following recipe:
\begin{itemize}
    \item[(i)] Standardize covariates so that $\sum_{i=1}^nX_{ij}=0$ and  $n^{-1}\sum_{i=1}^nX_{ij}^2=1$ for all $j$ (excluding the intercept)
    \item[(ii)] Set $Q_0^{-1} = \frac{b}{p+n}I_n$ for some fixed $b>0$, e.g.\ $b=10$
    \item[(iii)] If $X$ contains an intercept, sample from $\pi(\beta)$ using Algorithm \ref{alg:PDA2}, otherwise using Algorithm \ref{alg:PDA}.
\end{itemize}
Our results suggest that, if the design matrix $X$ is not too far from a random one as in Assumption  \ref{asmp:no_int} or \ref{asmp:with_int}, the above recipe leads to mixing times that remain bounded (and fairly small) for all ranges of $n$ and $p$, thus resulting in computational robustness and efficiency of $\PDA$ and $\PGS$.

\subsection{Open questions}
We now briefly discuss some questions and open problems arising from the above exploration:
\begin{enumerate}
\item Even if the above findings show that the mixing times of $\PDA$ and $\PGS$ remain bounded as $n,p\to\infty$ in various settings, they also identify situations where this does not happen. 
For example, the upper bound in Corollary \ref{crl:noint} suggests that the mixing times may increase as $n/p$ increases, when $Q_0 = cI_p$ with fixed $c$ and $X$ contains no intercept, which is what we observe in the simulations. Similarly, Proposition \ref{prop:lower_bound_KL}, and the corresponding empirical study, shows that the mixing time of $\PDA$ increases linearly with $n$ in the case of fully imbalanced data and presence of the intercept (see also \cite{johndrow2019mcmc}). 
While these issues can be solved as discussed in Section \ref{sec:take_aways}, this requires adapting the prior to $n$. 
It is thus natural to wonder whether it is possible to find a $\pi$-invariant Markov operator $P$ whose mixing times are provably bounded uniformly over $y$, $n$ and $p$ under Assumptions \ref{asmp:no_int} or \ref{asmp:with_int} with fixed $c$. For example, one might look for $P$ such that
\[
\inf_{y}\lim_{n\to \infty} \, 
\rho_{EC}(P, \pi)>0
\]
almost surely, both when $p$ is fixed and when it grows with $n$, where $\rho_{EC}$ is the entropy contraction coefficient defined in \eqref{eq:rho_ec_def}. For example, a good candidate for $P$ is given by the interweaving strategy proposed in \cite{yu2011center}, which alternates a centered and non-centered step. However our proof strategy is not applicable any more, since the results of \cite{ascolani2024entropy} do not apply for the non-centered step of the algorithm.
\item The results of Theorem \ref{theo:mixing_times} hold uniformly over $y$, which means that they are worst-case with respect to $y$. As we highlighted in the simulation studies of Section \ref{sec:simulation} the latter can significantly differ from the average case, which can be a reasonable assumption in many scenarios: it would be interesting to develop upper bounds that capture the dependence on $y$ and, e.g., differ for well-specified data as opposed to worst-case ones.
\item Finally, another open question is to analyze mixing times with other priors on $\beta$. In this paper we focused on the normal distribution (which corresponds to a ridge penalization), but in high-dimensional scenarios a sparsity inducing prior, like the spike and slab \citep{george1993variable} or Horseshoe \citep{carvalho2009handling}, might be preferred. While it would be of great interest to extend our results in this direction, our proof technique heavily relies on Gaussianity (and more generally log-concavity), see Section \ref{sec:main_results} for more details. Extensions to non log-concave priors will likely require different and novel mathematical tools.
\end{enumerate}

\section{General theory about entropy contraction of two-block Gibbs Samplers}\label{sec:theory}

In this section we provide some general results on the entropy contraction coefficients of two-block deterministic-scan Gibbs Samplers (aka Data Augmentation kernels), of which the kernel $\PDA$ defined in \eqref{eq:AC_algorithm} is a special case.
Since the results of this section apply more generally than model \eqref{eq:posterior} and can be of independent interest, we use a general notation defined as follows.

\subsection{Data augmentation and marginal kernels}
Let 
$\pi\in\mathcal{P}(\sX)$
with $\sX=\sX_1\times\sX_2$.
We denote by $\pi_i\in\mathcal{P}(\sX_i)$ its $i$-th marginal and by $\Pi_{i\to j}$ the conditional distributions of $x_j$ given $x_i$ under $\pi$, i.e.\ $\Pi_{i\to j}$ is a Markov kernel from $\sX_i$ to $\sX_j$ defined as 
$\Pi_{i\to j}(x_i,A_j)= 
\mathbb{P}_{(X_1,X_2)\sim \pi}(X_j\in A_j|X_i = x_i)$
for every $x_i\in\sX_i$ and $A_j\subseteq \sX_j$.
The two-block deterministic-scan Gibbs Sampler on $\pi$ is a Markov transition kernel on $\sX$ defined as 
$\PDA=P_2P_1$, where 
\begin{equation}
P_1(x, \d x') = \delta_{x_2}(\d x_2')\Pi_{2\to 1}(x_2,\d x_1),
\hbox{ and }
P_2(x, \d x') = \delta_{x_1}(\d x_1')\Pi_{1\to 2}(x_1,\d x_2)\,.
\end{equation}
If $\{(X_1^{(t)},X_2^{(t)})\}_t$ is a Markov chain with kernel $\PDA$, then $\{X_2^{(t)}\}_t$ is also a Markov chain and it has kernel $\PMG=\Pi_{1 \to 2}\Pi_{2 \to 1}$, see e.g.\ \cite[Section 3.3]{roberts2001markov}.
The convergence properties of $\PDA$ are closely related to the ones of $\PMG$, as shown in the following proposition.

\begin{proposition}\label{prop: equivalence_AC_MG}
Let $\pi,\mu \in \sP(\sX_1\times\sX_2)$ and $t \geq 1$. Then
\[
\KL\left(\mu \PDA^{t+1}, \pi \right) \leq \KL\left(\mu_2 \PMG^{t}, \pi_2 \right) \leq \KL\left(\mu \PDA^{t}, \pi \right)\,.
\]
\end{proposition}
\begin{proof}
By definition of $\PMG$, we have 
$\int_{\sX_1}\mu \PDA^t(\d x_1, A) = \mu_2\PMG^t(A)$, 
for every $A \subset \sX_2$. By the chain rule for the KL, this implies $\KL\left(\mu_2 \PMG^{t}, \pi_2 \right) \leq \KL\left(\mu \PDA^{t}, \pi \right)$. 
To prove the other inequality, consider the Markov kernel from $\sX_2$ to $\sX$ defined as $\Pi_{2 \to (1,2)}(x_2,\d x')=\Pi_{2 \to 1}(x_2,\d x_1')\Pi_{1\to 2}(x_1',\d x_2')$, so that $\mu\PDA^{t+1}=\mu_2\PMG^t\Pi_{2 \to (1,2)}$ for all $t\geq 0$. Combining the latter with $\pi_2\Pi_{2 \to (1,2)}=\pi$, and the chain rule we have 
\[
\KL\left(\mu \PDA^{t+1}, \pi \right)=
\KL\left(\mu_2\PMG^t\Pi_{2 \to (1,2)}, \pi_2\Pi_{2 \to (1,2)} \right)
\leq \KL\left(\mu_2\PMG^t, \pi_2\right),
\]
as desired.
\end{proof}

Proposition \ref{prop: equivalence_AC_MG} implies that $\tmix(\epsilon,\mu, \PDA) \leq 1+\tmix(\epsilon,\mu_2, \PMG)$, which allows us to focus on $\PMG$, which enjoys more analytical tractability. 

\subsection{Entropy contraction}\label{sec:EC_of_DA}
For a Markov kernel $P$ from $\sX$ to $\sY$ and a distribution $\nu\in\sP(\sX)$, define the entropy contraction coefficient of $P$ relative to $\nu$ as
\begin{align}\label{eq:rho_ec_def}
\rho_{EC}(P,\nu)=\sup_{\mu\in \mathcal{M}}
\frac{\KL(\mu P, \nu P)}{\KL(\mu, \nu)}\,.
\end{align}
where $\mathcal{M} = \left\{\mu \in \sP(\sX) \mid \KL(\mu, \nu) < \infty \right\}$. 
The next lemma shows that $\rho_{EC}$ is sub-multiplicative.
\begin{lemma}\label{lm:sub-multiplicative}
Let $P$ be a kernel from $\sX$ to $\sY$ and $Q$ a kernel from $\sY$ to $\sZ$. 
Then, for every $\nu \in \sP(\sX)$, we have
\[
\rho_{EC}(QP,\nu)\leq \rho_{EC}(P,\nu)\rho_{EC}(Q,\nu P)
\]
\end{lemma}
\begin{proof}
Fix $\nu \in \sP(\sX)$ and note that $\nu P \in \sP(\sY)$. For every $\mu \in \sP(\sX)$ we have $\mu (QP)= (\mu P)Q$  and
\[
\begin{aligned}
  \frac{\KL(\mu (QP), \nu (QP))}{\KL(\mu, \nu)}  & = \frac{\KL((\mu P)Q, (\nu P)Q)}{\KL(\mu, \nu)}\\
  &= \frac{\KL((\mu P)Q, (\nu P)Q)}{\KL(\mu P, \nu P)} \,\frac{\KL(\mu P, \nu P)}{\KL(\mu, \nu)} \leq \rho_{EC}(Q,\nu P)\rho_{EC}(P,\nu).
\end{aligned}
\]
The result follows since $\mu$ is arbitrary.
\end{proof}

Interestingly, the so-called approximate tensorization of the entropy for $\pi $, i.e.\ the inequality in \eqref{eq:functional_2d} below, controls both $\rho_{EC}(\Pi_{2\to 1})$ and $\rho_{EC}(\Pi_{1\to 2})$, as shown below.
\begin{theorem}\label{theo:DA_general}
Let $\sX=\sX_1\times\sX_2$ and $\pi \in \sP(\sX)$. 
If
\begin{align}\label{eq:functional_2d}
\frac{\KL(\mu_1 , \pi_1)+\KL(\mu_2 , \pi_2)}{2}
&\leq
\left( 1 - \frac{1}{2\kappa^*} \right) \KL(\mu , \pi)
&
\end{align}
for all $\mu\in\sP(\sX)$, then
\begin{align}\label{eq:contraction_product}
\max\{\rho_{EC}(\Pi_{2\to 1}, \pi_2),\rho_{EC}(\Pi_{1\to 2}, \pi_1)\}
\leq
\left( 1 - \frac{1}{\kappa^*} \right)
\end{align}
and $\rho_{EC}(\PMG,\pi_2)\leq (1 - 1/\kappa^*)^2$.
\end{theorem}
\begin{proof}
Let $\mu_1 \in \sP(\sX_1)$. Applying \eqref{eq:functional_2d} to the measure $\mu(\d x) =\mu_1(\d x_1) \Pi_{1\to2}(x_1, \d x_2)\in \sP(\sX)$ we obtain
\begin{align}\label{eq:bound_mu_composed}
\frac{\KL\left(\mu_1, \pi_1 \right)+\KL\left(\mu_1\Pi_{1\to2} , \pi_2 \right)}{2}
    \leq \left( 1 - \frac{1}{2\kappa^*} \right) \KL(\mu, \pi)\,.
\end{align}
Since $\KL(\mu , \pi) = \KL(\mu_1 , \pi_1)$, which follows by definition of $\mu$ and the chain rule for the KL, \eqref{eq:bound_mu_composed} can be written as
\[
\frac{\KL\left(\mu_1\Pi_{1\to2}, \pi_2 \right)}{2} \leq \left( 1 - \frac{1}{2\kappa^*}-\frac{1}{2} \right) \KL(\mu_1, \pi_1)
=\frac{1}{2}\left( 1 - \frac{1}{\kappa^*}\right) \KL(\mu_1, \pi_1),
\]
which, together with $\pi_1=\pi_2\Pi_{2\to 1}$, implies $\rho_{EC}(\Pi_{2\to 1}, \pi_2)\leq (1 - 1/\kappa^*)$. 
Also  
$\rho_{EC}(\Pi_{1\to2}, \pi_1)\leq (1 - 1/\kappa^*)$, and thus \eqref{eq:contraction_product}, follows by symmetry.
Finally $\rho_{EC}(\PMG,\pi_2)\leq (1 - 1/\kappa^*)^2$ follows from \eqref{eq:contraction_product} and $\PMG=\Pi_{1\to2}\Pi_{2\to 1}$ by Lemma \ref{lm:sub-multiplicative}.
\end{proof}
\begin{remark}
Recall that $\Pi_{1\to 2}$ and $\Pi_{2\to 1}$ are adjoint of each other with respect to the inner products of $L^2(\pi_1)$ and $L^2(\pi_2)$, and thus have the same operator norm in $L^2$, i.e.\ $\|\Pi_{1\to 2}\|_{L^2}=\|\Pi_{2\to 1}\|_{L^2}$.
Nonetheless, $\rho_{EC}(\Pi_{1\to 2})\neq \rho_{EC}(\Pi_{2\to 1})$  in general  and the ratio of the two can be arbitrarily large (see e.g.\ Example 16 in \cite{caputo2024entropy}).
More generally, the connection between the entropy contraction coefficients of $\Pi_{1\to 2}$, $\Pi_{2\to 1}$, $\PMG$ and $\PDA$ is more subtle than for their $L^2$-norms (or equivalently spectral Gaps), see \cite{caputo2024entropy} for a detailed review, as well as \citet{andrieu2016random, qin2022convergence, chlebicka2025solidarity} for results in the $L^2$ context.
\end{remark}


In Section \ref{sec:proof_main} of the Supplementary Material we combine the results of this section with the ones in \cite{ascolani2024entropy} to prove \eqref{eq:bound_mix_time_GS}. 
More generally, the proof of Theorem \ref{theo:mixing_times} relies on bounds of the form
\begin{equation}\label{eq:mix_rho_EC}
    \tmix(\epsilon,\mu,P)\leq \frac{1}{1-\rho_{EC}(P,\pi)}\log\left(\frac{\KL(\mu,\pi)}{\epsilon}\right)\,,
\end{equation}
which directly follows from $\KL(\mu P^t,\pi)\leq\rho_{EC}(P,\pi)^t \KL(\mu,\pi)$
and $\log(1-1/c) \leq -1/c$ for $c > 1$.

\section{Starting distribution}\label{sec:starting_distribution}
To be fully informative, the results of Theorem \ref{theo:mixing_times} require to find a starting distribution $\mu \in \sP(\R^n \times \R^p)$ such that $\log(\KL(\mu , \pi))$ can be suitably controlled, which is what we do on this section. 
Since sampling from $\pi(\d z \mid \beta)$ is feasible, 
we take starting distributions 
$\mu \in \sP(\R^p \times \R^n)$ of the form
\begin{equation}\label{eq:warm_start}
\mu(\d z,\d \beta) = \pi(\d z \mid \beta)\mu_2(\d \beta).
\end{equation}
for some $\mu_2 \in \sP(R^n)$, so that $\KL(\mu , \pi)=\KL(\mu_2, \pi_2)$.
In \cite[Sec.$3.2.1$]{lee2024fast} it is proven that a Gaussian distribution centered around the mode of $\pi(\beta)$, and with a suitable variance, is a feasible starting distribution with good control in KL. Here instead we assume to start from the prior, i.e.\ take $\mu_2(\beta) = N(\beta \mid m,Q_0^{-1})$. This is arguably an easier choice, which does not require additional computations to find the mode, nor knowledge of smoothness constant to tune variances. The next proposition shows that such starting distribution is also close enough in $KL$ to lead to good mixing times bounds. We consider the case of prior with zero mean for simplicity, even if the result could be generalized at the price of slightly more complicated bounds.

\begin{proposition}\label{prop:starting_distribution}
Let $\mu \in \sP(\R^p \times \R^n)$ be as in \eqref{eq:warm_start}
with $\mu_2(\beta) = N(\beta \mid m,Q_0^{-1})$. Let $m=(0,\dots,0)^T\in\R^p$. 
Then, for every $y$, we have
\[
\begin{aligned}
\log&(\KL(\mu , \pi))
\leq \log \left( 2n+
n\log\left(2(1+n\lambda_{max}(X^TQ_0^{-1}X)\right)\right)\,.
\end{aligned}
\]
\end{proposition}


\acks{GZ acknowledges support from the European Research Council (ERC), through the Starting Grant (StG) `PrSc-HDBayLe', project number 101076564.}


\vskip 0.2in
\bibliography{bibliography}

\newpage

\appendix

\section{Implementation and cost per iteration}\label{sec:cost_per_iteration}

We now discuss the computational cost associated to run a single iteration of $\PDA$ and $n$ iterations of $\PGS$, separating the case $n > p$ from the one $p > n$. For ease of notation, we will denote $V = (X^TX+Q_0)^{-1}$.

\paragraph{$n > p.$} In the case of $\PDA$ in \eqref{eq:AC_algorithm}, the main cost is associated to the computation of conditional mean and covariance matrix of $\beta$ in \eqref{eq:posterior_beta}. Since $n > p$, then $V = \text{Var}(\beta \mid z, y)$ can be pre-computed with $\sO(np^2)$ cost (the matrix multiplication dominates the $\sO(p^3)$ cost of inversion). With the same cost, both $VX^T$ and $VQ_0\mu$ can be pre-computed. Then, for every iteration, $\E[\beta \mid z, y] = V(Q_0\mu+X^Tz)$ and $\E[z \mid \beta] = X\beta$ can be computed with $\sO(np)$ cost. Thus, the overall cost is given by $\sO(np^2)$ pre-computation and $\sO(np)$ per iteration.

As regards $\PGS$, the conditional distribution of $z_i$ can be written (see e.g. \cite{held2006bayesian}) as
\[
\pi(z_i \mid z_{-i}) \, \propto \, N\left(z_i \mid (1-h_i)^{-1}x_i^TVX^T(z-Q_0m)-h_i(1-h_i)^{-1}z_i, (1-h_i)^{-1}\right)\mathbbm{1}(y_i = g(z_i)),
\]
where $h_i = x_i^TVx_i$. Thus, after pre-computing $VX^T$ with $\sO(np^2)$ cost, then $x_i^TV$ is given by the $i$-th column of $XV$. Then also $x_i^TVx_i$ can be pre-computed for every $i$, with overall cost $\sO(np)$. After updating the $i$-th component, the matrix $B = VX^T$ can be updated at $\sO(p)$ cost by noticing that
\[
B = B_{\text{old}}+S_i(z_i-z_{i,\text{old}}),
\]
where $S_i$ is the $i$-th column of $VX^T$, while $B_{\text{old}}$ and $z_{i,\text{old}}$ refer to the values before updating $z_i$.
Thus, the conditional mean can be obtained at $\sO(p)$. This implies that the overall cost is given by $\sO(np^2)$ pre-computation and $\sO(np)$ for every $n$ iterations. Finally, if needed, a sample of $\beta$ can be obtained with an additional $\sO(np)$ cost.

\paragraph{$p > n.$} As regards $\PDA$ we can pre-compute $V$ using the Woodbury's identity, which reads
\[
V = Q_0^{-1}-Q_0^{-1}X^T(I_n + XQ_0^{-1}X^T)^{-1}X,
\]
with a $\sO(n^2p)$ cost (since the $\sO(n^3)$ cost of inversion is dominated by the matrix multiplication). Similarly, it is possible to pre-compute
\[
\bar{V} = VX^T = Q_0^{-1}X^T-Q_0^{-1}X^T(I_n + XQ_0^{-1}X^T)^{-1}XX^T
\]
again at $\sO(n^2p)$ cost. Then, for every iteration it is possible to compute the conditional means $X\beta$ and $\bar{V}z$ at $\sO(np)$ cost. This implies that the overall cost is given by $\sO(n^2p)$ pre-computation and $\sO(np)$ for every iteration. 

As an alternative it is possible to implement instead the Markov chain with operator $\PDAtilde$, which samples from the full conditionals of $z$ and $\tilde{\beta} = X\beta$. The full conditionals are then given by $\tilde{\pi}(z \mid \tilde{\beta}) \, \propto \, N(z \mid \tilde{\beta}, I_n)\prod_{i = 1}^n\mathbbm{1}(y_i = g(z_i))$ and
\[
\tilde{\pi}(\tilde{\beta} \mid z) = N\left(\beta \mid X(X^TX+Q_0)^{-1}(Q_0m+X^Tz) , X(X^TX+Q_0)^{-1}X^T\right).
\]
By construction 
$\mu \PDA^t \in \sP(\R^p\times \R^n)$ and $\mu \PDAtilde^t \in \sP(\R^n\times \R^n)$, with 
$\int_{\R^p}\mu \PDA^t(A, \d \beta) = \int_{\R^n}\nu \PDAtilde^t(A, \d \tilde{\beta})$ for every $A \subset \R^n$, $\mu \in \sP(\R^n\times \R^p)$, and $\nu \in \sP(\R^n\times \R^n)$ with $\nu_2 = X\circ\mu_2$, i.e. the marginal distribution on $z$ is the same. Moreover they are co-deinitializing in the sense of \cite{roberts2001markov}, which implies that the two chains enjoy the same convergence properties. This is formalized in the next lemma.
\begin{lemma}
Fix $t \geq 1$ and let $\mu \in \sP(\R^n\times \R^p)$ and $\nu \in \sP(\R^n\times \R^n)$ with $\nu_2 = X\circ\mu_2$. Then we have that
\[
\KL(\mu \PDA^t, \pi) = \KL(\nu \PDAtilde^t, \tilde{\pi}).
\]
\end{lemma}
\begin{proof}
Consider $\tau \in \sP(\R^n)$ defined as $\tau(A) = \int_{\R^p}\mu \PDA^t(A, \d \beta) = \int_{\R^n}\nu \PDAtilde^t(A, \d \tilde{\beta})$, for every $A \subset \R^n$. By the chain rule for the KL divergence, we have that
\[
\KL(\mu \PDA^t, \pi) = \KL(\tau, \pi_1)+\E_{z \sim \tau}\left[\KL(\pi(\d \beta \mid z), \pi(\d \beta \mid z)\right] = \KL(\tau, \pi_1).
\]
The same argument holds for $\PDAtilde$.
\end{proof}
Using again Woodbury's identity, it is possible to compute $W = X(X^TX+Q_0)^{-1}X^T$ with $\sO(n^2p)$ operations. Then each iteration only requires to compute $Wz$, at $\sO(n^2)$ cost. This implies that the overall cost is given by $\sO(n^2p)$ pre-computation and $\sO(n^2)$ for every $n$ iterations. Of course, if needed, a sample of $\beta$ can be obtained with an additional $\sO(np)$ cost.

As regards $\PGS$, the prior precision matrix $Q = (I_n + XQ_0^{-1}X^T)^{-1}$ can be pre-computed at $\sO(n^2p)$ cost and the conditional distributions can be rewritten as
\[
\pi(z_i \mid z_{-i}, y) \, \propto \, N\left(z_i \mid -Q_{ii}^{-1}Q_{i,-i}^T(z_{-i}-(Xm)_{-i}), Q_{ii}^{-1}\right)\mathbbm{1}(y_i = g(z_i)),
\]
where $Q_{i,-i}$ denotes the $i$-th row of $Q$ without the $i$-th entry. Then every iteration can be performed at $\sO(n)$ cost. This implies that the overall cost is given by $\sO(n^2p)$ pre-computation and $\sO(n^2)$ for every $n$ iterations. Finally, if needed, a sample of $\beta$ can be obtained with an additional $\sO(np)$ cost.

\section{Model variations: tobit and cumulative probit regression}\label{sec:tobit}

As mentioned in Remark \ref{rmk:model_variations}, our results extend also to other models that can be expressed as partially or fully discretized Gaussian linear regression \citep{anceschi2023bayesian}. Here we consider two examples, which differ in the regression structure and choice of discretization mechanism.
The first example is the tobit model for censored data. Here the posterior, after some manipulations, can be traced back to the one of a probit model with modified data and priors, so that the mixing time bounds follow directly from Theorem \ref{theo:mixing_times}, but with a different prior matrix. The second one is the cumulative probit model for ordinal categorical data, whose posterior does not coincide with the one of probit models, but the same proof strategy of Theorem \ref{theo:mixing_times} applies analogously thanks to the convexity of the discretization sets. 
Other examples where Theorem \ref{theo:mixing_times} can be applied either directly or with minor variations include the classical and sequential multinomial probit models, 
see respectively Sections 2.1-2.2 and Section 2.3 of \citet{fasano2022class} for a self-contained description of such models. We do not discuss those here for brevity, since the overall message is analogous.

\paragraph{Tobit model}
The tobit model \citep{tobin1958estimation}
is a classical censored regression model where response data are observed only if they exceed a certain threshold, which is often set to $0$ for convenience. The resulting Bayesian model admits a natural data augmentation representation, analogous to \eqref{eq:probit_DA}, which reads
\begin{equation}\label{eq:tobit_DA}
\begin{aligned}
&y_i
= z_i\mathbbm{1}(z_i > 0)&i=1,\dots,n,\\
&z|\beta \sim N(X\beta,I_n), \quad \beta \sim N(m,Q_0^{-1})\,.
\end{aligned}
\end{equation}
Given the definition of $y_i$, it is convenient to partition the observations $y=(y_1,\dots,y_n)^T\in\R^n$ into censored ones, which we denote as
$\bar{y}=(\bar{y}_1,\dots,\bar{y}_{n_0})^T\in\R^{n_0}$ where $\bar{y}_i=0$ for all $i\in\{1,\dots,n_0\}$, and non-censored ones, which we denote as $\tilde{y}=(\tilde{y}_1,\dots,\tilde{y}_{n_1})^T\in\R^{n_1}$ where $\tilde{y}_i>0$ for all $i\in\{1,\dots,n_1\}$. Here $n_0+n_1=n$.
Similarly, we partition in an analogous way the latent variables $z=(z_1,\dots,z_n)^T\in\R^n$ into $\bar{z}=(\bar{z}_1,\dots,\bar{z}_{n_0})^T\in\R^{n_0}$
and $\tilde{z}=(\tilde{z}_1,\dots,\tilde{z}_{n_1})^T\in\R^{n_1}$, 
where $\bar{z}_i \leq 0$ for all $i\in\{1,\dots,n_0\}$ and  $\tilde{z}_i >0$ for all $i\in\{1,\dots,n_0\}$, and the design matrix $X$ into $\bar{X} \in \R^{n_0 \times p}$ and $\tilde{X} \in \R^{n_1 \times p}$. 
Given the deterministic constraint $\tilde{z}=\tilde{y}$, one needs only to consider the joint posterior of $\bar{z}$ and $\beta$, which reads
\begin{align}\label{eq:tobit_posterior}
\bar{\pi}(\bar{z}, \beta) \, &\propto \, 
N(\beta \mid m,Q_0^{-1})
N(\tilde{y} \mid \tilde X \beta,I_{n_1})
N(\bar{z} \mid \bar{X} \beta,I_{n_0})
\prod_{i = 1}^{n_0}\mathbbm{1}\left(\bar{y}_i = g(\bar{z}_i)\right)
\\
&\propto \, 
N\left(\beta \mid m_{new},Q_{new}^{-1}\right)
N(\bar{z} \mid \bar X \beta,I_{n_0})
\prod_{i = 1}^{n_0}\mathbbm{1}\left(\bar{y}_i = g(\bar{z}_i)\right),\label{eq:tobit_posterior_2}
\end{align}
with $g(\bar{z}_i) = \mathbbm{1}(\bar{z}_i > 0)$, $m_{new}=(\tilde{X}^T\tilde{X}+Q_0)^{-1}(Q_0m+\tilde{X}^T\tilde{y})$ and $Q_{new}=\tilde{X}^T\tilde{X}+Q_0$.
Also in the tobit context, posterior samples are often drawn using the corresponding data-augmentation algorithm \citep{chib1992bayes}, i.e.\ the two-block deterministic-scan Gibbs Sampler $\bar{P}_\mathrm{DA}$ targeting $\bar{\pi}(\bar{z}, \beta)$ by alternating the update of $\bar{z}$ from $\bar{\pi}(\bar{z} \mid \beta)$ and $\beta$ from $\bar{\pi}(\beta\mid \bar{z})$. Alternatively, one could consider the collapsed kernel $\bar{P}_\mathrm{CG}$, targeting directly the marginal distribution  $\bar{\pi}(\bar{z} )$.
Comparing \eqref{eq:tobit_posterior_2} with \eqref{eq:posterior} shows that $\bar{\pi}$ can be interpreted as the posterior distribution of a probit model with $n_0$ observations all equal to $0$, and a modified prior $\beta$ equal to $N(m_{new},Q_{new}^{-1})$.
It thus follows from Theorem \ref{theo:mixing_times} that the mixing times of $\bar{P}_\mathrm{DA}$ and $\bar{P}_\mathrm{CG}$ are upper bounded as 
\begin{align}
\tmix(\epsilon,\mu, \bar{P}_\mathrm{DA}) &\leq (2+\lambda_{max}(\bar{X}Q_{new}^{-1}\bar{X}^T))\log\left(\frac{\KL(\mu, \pi)}{\epsilon}\right)
\,,\\
\tmix(\epsilon,\mu_1, \bar{P}_\mathrm{CG}^{n_0}) &\leq \frac{1+\lambda_{max}(\bar{X}Q_{new}^{-1}\bar{X}^T)}{1+\lambda_{min}(\bar{X}Q_{new}^{-1}\bar{X}^T)}\log\left(\frac{\KL(\mu, \pi)}{\epsilon}\right)\,.
\end{align}
Interestingly, the spectrum of $\bar{X}Q_{new}^{-1}\bar{X}^T$ behaves differently than the one of 
$XQ_0^{-1}X^T$. 
For example, under random design assumptions, the quantity $\lambda_{max}(XQ_0^{-1}X^T)$ diverges with the ratio $r\approx n/p$ of number of data points over parameters (see Corollary \ref{crl:noint}), while for $\lambda_{max}(\bar{X}Q_{new}^{-1}\bar{X}^T)$ to remain bounded it is sufficient that the ratio $\zeta\approx n_0/n_1$ of censored versus uncensored observations does so, as shown in the following corollary.

\begin{corollary}\label{crl:tobit}
Under Assumption \ref{asmp:no_int} we have that
\[
\tmix(\epsilon,\mu, \bar{P}_\mathrm{DA}) \leq (2+\zeta+\delta)\log\left(\KL(\mu,\pi)/\epsilon\right)
\]
and
\[
\tmix(\epsilon,\mu_1, \bar{P}_\mathrm{CG}^{n_0}) \leq 
(2+\zeta+\delta)\log\left(\KL(\mu,\pi)/\epsilon\right),
\]
almost surely as $n_0,n_1 \to \infty$ with $p$ fixed and $n_0/n_1 \to \zeta \in (0, \infty)$, for any arbitrarily small positive constant $\delta > 0$.
\end{corollary}
\begin{proof}
By definition of $Q_{new}$ and standard matrix properties, we have
\begin{align*}
\lambda_{max}(\bar{X}Q_{new}^{-1}\bar{X}^T)
=
\lambda_{max}(\bar{X}(\tilde{X}^T\tilde{X}+Q_0)^{-1}\bar{X}^T)
=
\lambda_{max}((\tilde{X}^T\tilde{X}+Q_0)^{-1}\bar{X}^T\bar{X})\,.
\end{align*}
Since both $\tilde{X}^T\tilde{X}+Q_0$ and $\bar{X}^T\bar{X}$ have fixed dimensionality as $n_0,n_1 \to \infty$ with $p$ fixed and $n_0/n_1 \to \zeta \in (0, \infty)$, by Assumption \ref{asmp:no_int}, the strong law of large numbers and continuity of the matrix inverse, we have
\begin{align}\label{eq:tobit_fixed_d_lim}
(\tilde{X}^T\tilde{X}+Q_0)^{-1}\bar{X}^T\bar{X}
=
\frac{n_0}{n_1}
\left(\frac{\tilde{X}^T\tilde{X}+Q_0}{n_1}\right)^{-1}
\frac{\bar{X}^T\bar{X}}{n_0}
\to \zeta I_p    
\end{align}
almost surely. 
Thus, also $\lambda_{max}(\bar{X}Q_{new}^{-1}\bar{X}^T)$ converges almost surely to $\zeta$  by continuity of the operator $\lambda_{max}$.  
Instead, $\lambda_{min}(\bar{X}Q_{new}^{-1}\bar{X}^T)
= 0$ as soon as $n_0>p$ because $\bar{X} \in \R^{n_0 \times p}$ and thus the rank of $\bar{X}Q_{new}^{-1}\bar{X}^T$ is at most $p$.
The desired results then follow by combining  $\lambda_{max}(\bar{X}Q_{new}^{-1}\bar{X}^T)\to\gamma$ and $\lambda_{min}(\bar{X}Q_{new}^{-1}\bar{X}^T)\to 0$ with Theorem \ref{theo:mixing_times}.
\end{proof}

\paragraph{Cumulative probit model}
The cumulative probit (or ordered probit) model \citep{mccullagh1980regression} is a popular model for to perform regression with \emph{ordinal} categorical data \citep{agresti2010analysis}, such as responses of questionnaires measuring levels of agreement or datasets measuring severity of diseases in a discretized scale. Similarly to \eqref{eq:probit_DA} and \eqref{eq:tobit_DA}, the cumulative probit model assumes that each ordinal observation $y_i\in\{1,\dots,K\}$, where $K$ denotes the number of categories, arises as the discretization of some continuous latent variable $z_i$. Specifically, the model reads
\begin{equation}\label{eq:cumulative_probit_DA}
\begin{aligned}
&y_i
=g_\alpha(z_i)=\sum_{k=1}^K
k\mathbbm{1}(\alpha_{k-1}< z_i \leq \alpha_k)
&i=1,\dots,n,\\
&z|\beta \sim N(X\beta,I_n), \quad \beta \sim N(m,Q_0^{-1})\,,
\end{aligned}
\end{equation}
with $-\infty=\alpha_0<\alpha_1<\dots<\alpha_K=\infty$ being a set of real-valued thresholds. Here we assume $(\alpha_k)_{k=1}^K$ to be fixed for simplicity, and refer to \cite{aliverti2025approximate} for more discussion of how these thresholds are tuned or estimated in practice. 
The joint posterior density of $z$ and $\beta$ coincides with \eqref{eq:posterior}, but with $g$ replaced by $g_\alpha$, i.e.\ it reads
\begin{equation}\label{eq:posterior_cumulative}
\pi_\alpha(z, \beta) \, \propto \, N(\beta \mid m,Q_0^{-1})N(z \mid X\beta,I_n)\prod_{i = 1}^n\mathbbm{1}\left(y_i = g_\alpha(z_i)\right)\,.
\end{equation}
It follows that the conditional density $\pi_\alpha(\beta \mid z)$ coincide with $\pi(\beta \mid z)$ defined in \eqref{eq:posterior_beta}, while $\pi_\alpha(z \mid \beta)$ reads
$$\pi_\alpha(z \mid \beta) \, \propto \, N(z \mid X\beta, I_n)\prod_{i = 1}^n\mathbbm{1}(\alpha_{y_i-1}<z_i\leq \alpha_{y_i})\,,$$
which factorizes as a product of one-dimensional Gaussians truncated on an interval, thus being easy to sample from.

Crucially, the sets 
$A_{y_i}^{(\alpha)}:=\{z_i\in\R\,:\,y_i = g_\alpha(z_i)\}$ appearing in \eqref{eq:posterior_cumulative} are convex, like the analogous sets 
$A_{y_i}:=\{z_i\in\R\,:\,\mathbbm{1}\left(y_i = g(z_i)\right)\}$ in \eqref{eq:posterior}.
Thus, the proof of Theorem \ref{theo:mixing_times} in Section \ref{sec:proof_main} applies analogously\footnote{Specifically, the only difference relative to the proof in Section \ref{sec:proof_main} is to replace the approximating functions $U_{i, N}(z_i)
=
N\,\hbox{dist}(z_i,A_{y_i})$ 
with $U_{i, N}^{(\alpha)}(z_i)=N\,\hbox{dist}(z_i,A_{y_i}^{(\alpha)})$, 
where $\hbox{dist}(z,A)=\inf_{z'\in A}|z-z'|$ denotes the distance between a point $z\in\R$ and a set $A\subseteq \R$. 
In both cases, the resulting functions $U_{i,N}$ and $U_{i, N}^{(\alpha)}$ are convex and increasing in $N$, which allows to apply Proposition \ref{prop:appr_tensorization} as detailed in Section \ref{sec:proof_main}.
},
and the bounds in \eqref{eq:bound_mix_time_GS} and \eqref{eq:bound_mix_time_PCG} hold unchanged also in this context. Namely, for every $\mu \in \sP(\R^n\times \R^p)$ and $\epsilon>0$, we have
\begin{align*}
\tmix(\epsilon,\mu, \bar{P}^{(\alpha)}_\mathrm{DA}) &\leq (2+\lambda_{max}(XQ_0^{-1}X^T))\log\left(\frac{\KL(\mu, \pi)}{\epsilon}\right)\,,\\
\tmix(\epsilon,\mu_1, (\bar{P}^{(\alpha)}_\mathrm{CG})^n) &\leq \frac{1+\lambda_{max}(XQ_0^{-1}X^T)}{1+\lambda_{min}(XQ_0^{-1}X^T)}\log\left(\frac{\KL(\mu, \pi)}{\epsilon}\right)\,,
\end{align*}
where $\bar{P}^{(\alpha)}_\mathrm{DA}$ is the two-block deterministic-scan Gibbs Sampler targeting $\pi_\alpha(z, \beta)$ by alternating the update of $z$ from $\pi_\alpha(z\mid \beta)$ and $\beta$ from $\pi_\alpha(\beta\mid z)$, and $\bar{P}^{(\alpha)}_\mathrm{CG}$ is the $n$-block random-scan Gibbs Sampler defined as in \eqref{eq:GS_algorithm} but with $\pi$ replaced by $\pi_\alpha$.


\section{Additional useful results}
\subsection{Condition number of \eqref{eq:model}}\label{sec:condition_number}
Given $\pi(\beta)$ as in \eqref{eq:model} with $m=(0,\dots,0)^T\in\R^p$, we can write
\begin{align}\label{eq:pot_def}
U(\beta)&= -\log(\pi(\beta)) =\frac{\beta^TQ_0\beta}{2}+\sum_{i=1}^n h\left(\text{sgn}(2y_i-1)x_i^T\beta\right), \quad h(r) = -\log\left(\Phi(r)\right),
\end{align}
where $\text{sgn(u)}$ is the sign of $u \in \R$. Let $\tilde{U}$ be the prior-preconditioned version of $U$, i.e.\ $\tilde{U}(\theta)= U(Q_0^{-1/2}\theta)$ for $\theta\in\R^p$. We define the condition number of $\tilde{U}$ as
\[
\kappa(\tilde{U}):=\frac{\sup_{\theta\in\R^p}\lambda_{max}(\nabla^2\tilde{U}(\theta))}{\inf_{\theta\in\R^p}\lambda_{min}(\nabla^2\tilde{U}(\theta))}.
\]
We have a preliminary lemma.
\begin{lemma}\label{lm:technical_cond}
For every $r \in \R$ we have that $h''(r) \in (0, 1)$.
\end{lemma}
\begin{proof}
It is well-known that $\Phi(r)$ is a strictly log-concave function, i.e.\ that $h''(r) > 0$. Thus we focus on the upper bound. 
By simple calculations we get
\begin{align}\label{eq:h_curv_1}
h''(r) = \left(\frac{\phi(r)}{\Phi(r)} \right)^2+r\frac{\phi(r)}{\Phi(r)},
\end{align}
where $\phi(r) = N(r \mid 0, 1)$. Moreover, if $Z \sim N(0,1)$, it is easy to show
\begin{align}\label{eq:h_curv}
    0 \leq \text{Var}(Z \mid Z < r) = 1-\left(\frac{\phi(r)}{\Phi(r)} \right)^2-r\frac{\phi(r)}{\Phi(r)} = 1-h''(r),
\end{align}
from which we deduce that $h''(r) < 1.$
\end{proof}
The next proposition provides an upper bound on $\kappa(\tilde{U})$.
\begin{proposition}\label{prop:cond_number}
It holds that
$\kappa(\tilde{U})\leq 1+\lambda_{max}(XQ_0^{-1}X^T)$.
\end{proposition}
\begin{proof}
The Hessian of $\tilde U$ is 
\begin{align}
\nabla^2\tilde{U}(\theta)&=I_p+Q_0^{-1/2}X^TD(\theta)XQ_0^{-1/2},
\end{align}
with $D(\theta)$ diagonal and $D_{ii}(\theta)=h''\left(\text{sgn}(2y_i-1)x_i^T\beta\right)$. By Lemma \ref{lm:technical_cond} we deduce that $I_p \succeq D(\theta)\succeq 0$, which implies
\[
I_p+Q_0^{-1/2}X^TXQ_0^{-1/2} \succeq \nabla^2\tilde{U}(\theta) \succeq I_p.
\]
This implies that $\lambda_{min}\left( \nabla^2\tilde{U}(\theta)\right) \geq 1$ and
\[
\lambda_{max}\left( \nabla^2\tilde{U}(\theta)\right) \leq 1 + \lambda_{max}\left(Q_0^{-1/2}X^TXQ_0^{-1/2} \right) = 1+\lambda_{max}(XQ_0^{-1}X^T),
\]
as desired.
\end{proof}

\subsection{Approximate tensorization by convex approximations}
In this section we adapt the results in \cite{ascolani2024entropy} to accommodate for the presence of indicator functions in the target distribution of a Gibbs sampler.

Let $\sX = \R^d$ and consider a partition of $\sX$ in $M$ blocks with length $d_m$, i.e.\ $\sX = \bigtimes_{i = m}^M \sX_m$ with $\sX_m = \R^{d_m}$ for $m=1, \ldots, M$ and $d = d_1 + \ldots + d_M$. For a point $x=(x_1, \ldots, x_M)\in \R^d$, we write $x_{-m} = (x_1, \ldots, x_{m-1}, x_{m+1}, \ldots, x_M)$, which is an element of $\sX_{-m} = \bigtimes_{i \neq m} \sX_i$. Similarly, $\pi_{-m}$ denotes the marginal distribution of $\pi \in \sP(\sX)$ over $\sX_{-m}$.

Define $\pi \in \sP(\sX)$ with density
\begin{equation}\label{eq:pi_app}
    \pi(x) \, \propto \, \pi_0(x)e^{-\sum_{m = 1}^MU_m(x_m)},
\end{equation}
where $\pi_0(x) = N(x \mid m, Q^{-1})$ and $U_m:\sX_m\to \R\cup\{+\infty\}$. Let $L_m > 0$ be such that $Q_{mm}- L_m I_{d_m}$ is positive semi-definite, where $Q_{mm}$ is the $d_m \times d_m$ diagonal block of $Q$. Define
\begin{equation}\label{eq:kappa_gaussian}
\kappa^* = \frac{1}{\lambda_{min}\left( D^{-1/2}QD^{-1/2}\right)},
\end{equation}
where $D$ denotes the diagonal matrix with diagonal coefficients $L_1, L_2, \ldots, L_M$, with each $L_m$ repeated $d_m$ times, that is, $D_{mm} = L_m \Id_{d_m}$. The next proposition proves an approximate tensorization in terms of $\kappa^*$.
\begin{proposition}
\label{prop:appr_tensorization}
Let $\pi$ be as in \eqref{eq:pi_app} and assume there exists $\{\pi_N\}_N \subset \sP(\sX)$ defined as
\[
\pi_N(x) \, \propto \, \pi_0(x)e^{-\sum_{m = 1}^MU_{m, N}(x_m)}
\]
such that 
\begin{enumerate}
    \item $U_{m, N}$ is convex for every $m$ and $N$.
    \item $U_{m, N}(x_m) \to U_{m}(x_m)$ as $N \to \infty$ for every $m$ and $x_m$.
    \item $U_{m, N}(x_m)$ is increasing in $N$ for every $m$ and $x_m$.
\end{enumerate}
Then for any $\mu \in \sP(\R^d)$
\[
\frac{1}{M} \sum_{m=1}^M \KL(\mu_{-m} , \pi_{-m}) \leq \left( 1 - \frac{1}{\kappa^* M} \right) \KL(\mu , \pi),
\]
with $\kappa^*$ as in \eqref{eq:kappa_gaussian}.
\end{proposition}
\begin{proof}
By $2.$ we have that $\pi_N \to \pi$ weakly as $N \to \infty$. Moreover, since $U_{m, N}$ is convex, $\pi_N$ satisfies Assumption $B$ in \cite{ascolani2024entropy}. Thus, by lower semi-continuity of the KL and Theorem $3.1$ in \cite{ascolani2024entropy} we have
\[
\frac{1}{M} \sum_{m=1}^M \KL(\mu_{-m} , \pi_{-m}) \leq \frac{1}{M} \sum_{m=1}^M \lim \inf_{N \to \infty}\KL(\mu_{-m} , \pi_{N,-m}) \leq \left( 1 - \frac{1}{\kappa^* M} \right) \lim \inf_{N \to \infty}\KL(\mu , \pi_N).
\]
Notice that
\[
\begin{aligned}
\KL(\mu , \pi_N) =& \log \left(\int_{\R^d}\pi_0(x)e^{-\sum_{m = 1}^MU_{m, N}(x_m)} \, \d x \right) + \int_{\R^d}\log\left(\frac{\mu(x)}{\pi_0(x)} \right)\mu(\d x)\\
& + \sum_{m = 1}^M\int_{d_m}U_{m, N}(x_m)\mu_{m}(\d x_m).
\end{aligned}
\]
By dominated convergence theorem
\[
\int_{\R^d}\pi_0(x)e^{-\sum_{m = 1}^MU_{m, N}(x_m)} \, \d x \to \int_{\R^d}\pi_0(x)e^{-\sum_{m = 1}^MU_{m}(x_m)} \, \d x
\]
and by monotone convergence theorem (which holds by $3.$)
\[
\int_{\R^{d_m}}U_{m, N}(x_m)\mu_{m}(\d x_m) \to \int_{\R^{d_m}}U_{m}(x_m)\mu_{m}(\d x_m)
\]
for every $m = 1, \dots, M$ as $N \to \infty$. Thus we conclude that $\lim \inf_{N \to \infty}\KL(\mu , \pi_N) = \KL(\mu , \pi)$ from which the result follows.
\end{proof}
Let now $P$ be the transition kernel associated to the random scan Gibbs sampler on $\pi$ which alternates sampling from $x_m$ given $x_{-m}$.
\begin{corollary}
\label{crl:contraction_KL_appr}
Consider the same assumptions of Proposition \ref{prop:appr_tensorization}. Then for any $\mu \in \sP(\R^d)$
\[
\KL( \mu P, \pi ) \leq \left( 1 - \frac{1}{\kappa^* M} \right) \KL(\mu , \pi). 
\]
\end{corollary}
\begin{proof}
The proof is analogous to Theorem $3.2$ in \cite{ascolani2024entropy}, replacing Theorem $3.1$ therein with Proposition \ref{prop:appr_tensorization}.
\end{proof}
\subsection{Mixing times bounds in $\chi^2$}\label{sec:mixing_times_chi}
We first need a preliminary lemma.
\begin{lemma}\label{lm: equivalence_AC_MG_chisquare}
Let $\pi,\mu \in \sP(\sX_1\times\sX_2)$ and $t \geq 1$. Then
\[
 \chi^2\left(\mu \PDA^{t+1}, \pi \right) \leq \chi^2\left(\mu_2 \PMG^{t}, \pi_2 \right) \leq \chi^2\left(\mu \PDA^{t}, \pi \right)\,.
\]
In particular this implies 
\[
\tau_{\mathrm{mix},2}(\epsilon,\mu_2, \PMG) \leq \tau_{\mathrm{mix},2}(\epsilon,\mu, \PDA) \leq 1+\tau_{\mathrm{mix},2}(\epsilon,\mu_2, \PMG)\,.
\]
\end{lemma}
\begin{proof}
The proof is similar to the one of Proposition \ref{prop: equivalence_AC_MG}. As regards the lower bound, consider the Markov kernel from $\sX_2$ to $\sX$ defined as $\Pi_{2 \to (1,2)}(x_2,\d x')=\Pi_{2 \to 1}(x_2,\d x_1')\Pi_{1\to 2}(x_1',\d x_2')$, so that $\mu\PDA^{t+1}=\mu_2\PMG^t\Pi_{2 \to (1,2)}$ for all $t\geq 0$. Combining the latter with $\pi_2\Pi_{2 \to (1,2)}=\pi$, and the monotonicity of $\chi^2$ divergence we have that
\[
\chi^2\left(\mu \PDA^{t+1}, \pi \right)=
\chi^2\left(\mu_2\PMG^t\Pi_{2 \to (1,2)}, \pi_2\Pi_{2 \to (1,2)} \right)
\leq \chi^2\left(\mu_2\PMG^t, \pi_2\right),
\]
as desired.

As regards the upper bound, notice that
\[
\chi^2( \mu, \pi) = \int\left(\frac{\d \mu}{\d \pi}(x) \right)^2 \pi(\d x) - 1,
\]
so it suffices to show that
\begin{equation}\label{eq:chi_square_to_show}
\int_{\sX_2}\left(\frac{\d \mu_2\PMG^t}{\d \pi_2}(x_2) \right)^2 \pi_2(\d x_2) \leq \int_{\sX_2}\int_{\sX_1}\left(\frac{\d \mu\PDA^t}{\d \pi}(x) \right)^2 \pi(\d x)
\end{equation}
By definition of Radon-Nykodim derivative we have that
\begin{equation}\label{eq:RN1}
\mu_2\PMG^t(A) = \int_{A}\frac{\d \mu_2\PMG^t}{\d \pi_2}(x_2)\pi_2(\d x_2),
\end{equation}
for every $A \subset \sX_2$. Moreover, by definition of $\PMG$, we have 
$\int_{\sX_1}\mu \PDA^t(\d x_1, A) = \mu_2\PMG^t(A)$, 
which means
\begin{equation}\label{eq:RN2}
\begin{aligned}
\mu_2\PMG^t(A) &= \mu \PDA^t(\sX_1, A) = \int_{A} \int_{\sX_1}\frac{\d \mu\PDA^t}{\d \pi}(x_1, x_2) \pi(\d x_1, \d x_2)\\
& = \int_{A} \left[ \int_{\sX_1}\frac{\d \mu\PDA^t}{\d \pi}(x_1, x_2)\pi(\d x_1 \mid x_2)\right] \pi_2(\d x_2).
\end{aligned}
\end{equation}
Combining \eqref{eq:RN1} and \eqref{eq:RN2} we get that
\[
\frac{\d \mu_2\PMG^t}{\d \pi_2}(x_2) = \int_{\sX_1}\frac{\d \mu\PDA^t}{\d \pi}(x_1, x_2)\pi(\d x_1 \mid x_2),
\]
which means
\[
\begin{aligned}
    \int_{\sX_2}\left(\frac{\d \mu_2\PMG^t}{\d \pi_2}(x_2) \right)^2 \pi_2(\d x_2) & = \int_{\sX_2}\left(\int_{\sX_1}\frac{\d \mu\PDA^t}{\d \pi}(x_1, x_2)\pi(\d x_1 \mid x_2) \right)^2 \pi_2(\d x_2)\\
    & \leq \int_{\sX_2}\int_{\sX_1}\left(\frac{\d \mu\PDA^t}{\d \pi}(x_1, x_2)\right)^2\pi(\d x_1 \mid x_2)  \pi_2(\d x_2)\\
    & = \int_{\sX_2}\int_{\sX_1}\left(\frac{\d \mu\PDA^t}{\d \pi}(x)\right)^2\pi(x),
\end{aligned}
\]
and therefore \eqref{eq:chi_square_to_show} is proved.
\end{proof}
Denote the spectral gap of a $\pi$-reversible Markov kernel $P$ on $\sX$ as
\begin{align}\label{eq:gap}
    \gap(P) = \inf_f \, \frac{\int \int (f(y)-f(x))^2P(x, \d y) \pi(\d x)}{2\text{Var}_\pi(f)}
\end{align}
with the infimum running on every $f$ such that $\text{Var}_\pi(f) < \infty$. It is known \citep[below Lemma $2.15$]{caputo2023lecture} that
\begin{equation}\label{eq:general_lower_bound_gap}
\gap(P) \geq \frac{1-\rho_{EC}(P, \pi)}{2} \, ,
\end{equation}
and \citep[equation (5)]{AL22} that
\[
 \chi^2\left(\mu P^t, \pi \right)   \leq \left( 1-\gap(P)\right)^t\chi^2\left(\mu, \pi \right)\,.
\]
Combining the above equations we obtain that
\begin{equation}\label{eq:mix_rho_EC_chi}
    \tmixchi(\epsilon,\mu,P)\leq \frac{2}{1-\rho_{EC}(P,\pi)}\log\left(\frac{\chi^2(\mu,\pi)}{\epsilon}\right)\,.
\end{equation}
We can now prove the analogue of Theorem \ref{theo:mixing_times} for $\tmixchi$.
\begin{theorem}\label{theo:mixing_times_chi}
For every $\mu \in \sP(\R^n\times \R^p)$ and $\epsilon>0$, we have
\[
\tmixchi(\epsilon,\mu, \PDA) \leq (3+2\lambda_{max}(XQ_0^{-1}X^T))\log\left(\frac{\chi^2(\mu, \pi)}{\epsilon}\right)\,,
\]
and
\[
\tmixchi(\epsilon,\mu_1, \PGS^n) \leq 2\,\frac{1+\lambda_{max}(XQ_0^{-1}X^T)}{1+\lambda_{min}(XQ_0^{-1}X^T)}\log\left(\frac{\chi^2(\mu, \pi)}{\epsilon}\right)\,.
\]
\end{theorem}
\begin{proof}
By \eqref{eq:upper_bound_rho_PDA} below and \eqref{eq:mix_rho_EC_chi} we have that
\[
\tmixchi(\epsilon,\mu,\PMG)\leq \left( 2 + 2\lambda_{max}\left(XQ_0^{-1}X^T \right)\right)\log\left(\frac{\chi^2(\mu,\pi)}{\epsilon}\right)\,.
\]
Therefore the result for $\PDA$ follows by Lemma \ref{lm: equivalence_AC_MG_chisquare}. The result for $\PGS$ follows by combining \eqref{eq:upper_bound_rho_PCG} below and again \eqref{eq:mix_rho_EC_chi}.
\end{proof}

\section{Proof of the results in the main document}\label{sec:proofs_appendix}

\subsection{Proof of Theorem \ref{theo:mixing_times}}\label{sec:proof_main}
We first prove \eqref{eq:bound_mix_time_GS}.
\begin{proof}
We consider a re-parametrized version of model \eqref{eq:probit_DA}, where $\beta$ is replaced by 
$\tilde{\beta} = R \beta$ with $R=(Q_0+X^TX)^{1/2}$ and $R$ symmetric. Note that $R$ is always well defined since $Q_0$ is positive definite.
The model now reads
\begin{equation}\label{eq:transf_model}
\tilde{\beta} \sim N(\tilde{\beta} \mid Rm,RQ_0^{-1}R),
\quad
z|\tilde{\beta} \sim N(z \mid XR^{-1}\tilde{\beta},I_n), 
\quad 
y_i= \mathbbm{1}(z_i > 0)\hbox{ for }i=1,\dots,n\,,
\end{equation}
The joint posterior of $(z, \tilde{\beta})$ given $y$ under \eqref{eq:transf_model} is
\begin{equation}\label{eq:transf_post}
\tilde\pi( z, \tilde{\beta}) \, \propto \, N\left( (z, \tilde{\beta}) \mid \tilde{\mu}, \tilde{Q}^{-1}\right)\prod_{i = 1}^n\mathbbm{1}\left(y_i = g(z_i)\right),
\end{equation}
where $\tilde{\mu}=[(Xm)^T,(Rm)^T]^T$ and, since $R^{-1}(Q_0+X^TX)R^{-1}=I_p$,
\begin{equation}\label{eq:standard_precision}
  \tilde{Q}
=
\left( {\begin{array}{cc}
I_n & -XR^{-1} \\
-R^{-1}X^T & I_p\\
  \end{array} } \right)\,.
\end{equation}
Since $R$ is invertible, the two-block GS targeting $\pi$ and $\tilde{\pi}$ has exactly the same mixing times in KL (see e.g.\ \cite[Remark 2.3]{ascolani2024entropy} for details).

Thus, $\pi(z, \tilde{\beta}) = \lim_{N \to \infty}\pi_N(z, \tilde{\beta})$ with 
\[
\pi_N(z, \tilde{\beta}) \, \propto \,  N( (z, \tilde{\beta}) \mid \tilde{\mu}, \tilde{Q}^{-1})e^{-U_N(z)}
\]
and $U_N(z) = \sum_{i = 1}^nU_{i, N}(z_i)$ defined as
$U_{i, N}(z_i)=N|z_i|\mathbbm{1}\left(y_i \neq g(z_i)\right)$. 
Thus $\pi$ satisfies the assumptions in Proposition \ref{prop:appr_tensorization} with $M = 2$ and then
\[
\frac{\KL(\mu_1, \pi_1)+\KL(\mu_2, \pi_2)}{2} \leq
\left( 1 - \frac{1}{2\kappa^*} \right) \KL(\mu, \pi),
\]
for every $\mu \in \sP(\R^n \times \R^p)$, where $\kappa^* = 1/\lambda_{min}(\tilde{Q})$ (since $\tilde{Q}$ has identity block-diagonal terms).
Moreover, by standard linear algebra calculations \eqref{eq:standard_precision} implies 
\[
\lambda_{min}(\tilde{Q}) 
= 1-\sqrt{\lambda_{max}(XR^{-2}X^T)}
 = 1-\sqrt{\lambda_{max}(X(Q_0+X^TX)^{-1}X^T)}.
\]
see e.g.\ Lemma 2 in \cite{goplerud2024partially}. Therefore, by Theorem \ref{theo:DA_general} we have 
$\rho_{EC}(\PMG,\pi_2)\leq \lambda_{max}(X(Q_0+X^TX)^{-1}X^T)$.
Applying Woodbury's matrix identity twice, we obtain
\begin{align*}
X(Q_0+X^TX)^{-1}X^T 
&= 
X(Q_0^{-1}-Q_0^{-1}X^T
(I_n+XQ_0^{-1}X^T)^{-1}
XQ_0^{-1})X^T
\\
&=
M-M(I_p+M)^{-1}M
=(I+M^{-1})^{-1}    
\end{align*}
for $M=XQ_0^{-1}X^T$.
Thus 
\begin{equation}\label{eq:upper_bound_rho_PDA}
\begin{aligned}
\rho_{EC}(\PMG,\pi_2)
&\leq
\frac{1}{\lambda_{min}(I_n+M^{-1})}
&=
\frac{1}{1+1/\lambda_{max}(M)} 
= \frac{\lambda_{max}(M)}{1+\lambda_{max}(M)}\,.
\end{aligned}
\end{equation}
The result then follows from \eqref{eq:mix_rho_EC} and $\tmix(\epsilon,\mu, \PDA) \leq 1+\tmix(\epsilon,\mu_2, \PMG)$.
\end{proof}

The inequality in \eqref{eq:bound_mix_time_PCG} follows instead by the next theorem.
\begin{theorem}\label{theo:contraction_GS}
Let $\pi$ be as in \eqref{eq:posterior}, $\PGS$ as in \eqref{eq:GS_algorithm} and $\mu_1 \in \sP(\R^n)$.
Then
\begin{equation}\label{eq:mixing_times_GS}
\begin{aligned}
\tmix(\epsilon,\mu_1, \PGS^n)
&\leq
\lambda_{max}(D^{1/2}(I_n+M)D^{1/2})\log\left(\frac{\KL(\mu_1,\pi_1)}{\epsilon}\right)\\
&\leq
\left(\frac{1+\lambda_{max}(M)}{1+\lambda_{min}(M)}\right)\log\left(\frac{\KL(\mu_1,\pi_1)}{\epsilon}\right)\\
& \leq \left(1+\lambda_{max}(M)\right)\log\left(\frac{\KL(\mu_1,\pi_1)}{\epsilon}\right)\,,
\end{aligned}
\end{equation}
with $M=XQ_0^{-1}X^T$ and $D$ being a diagonal matrix with diagonal elements equal to the ones of $(I_n+M)^{-1}$.
\end{theorem}
\begin{proof}
The marginal distribution of $z$ under  \eqref{eq:posterior} is 
\[
\pi(z) \, \propto \, N(z \mid Xm,I_n + M)\prod_{i = 1}^n\mathbbm{1}\left(y_i = g(z_i)\right)\,.
\]
Thus, $\pi(z) = \lim_{N \to \infty}\pi_N(z)$ with 
$
\pi_N(z) \, \propto \, N(z \mid X\mu, I_n + M)e^{-\sum_{i = 1}^nU_{i, N}(z_i)}
$ 
and
$U_{i, N}(z_i)=N|z_i|\mathbbm{1}\left(y_i \neq g(z_i)\right)$ as above. 
Thus $\pi(z)$ satisfies the assumptions in Corollary \ref{crl:contraction_KL_appr}, implying
$\rho_{EC}(\PGS,\pi_1)\leq 1-1/(n\kappa^*)$
with
\[
\kappa^* = 1/\lambda_{min}\left(D^{-1/2}(I_n+M)^{-1}D^{-1/2}\right)
=
\lambda_{max}\left(D^{1/2}(I_n+M)D^{1/2}\right)\,.
\]
Combining the latter with \eqref{eq:mix_rho_EC} gives the first inequality in  \eqref{eq:mixing_times_GS}.  
Then, by Lemma $2.4$ in \cite{ascolani2024entropy} we have that
\[
\begin{aligned}
\kappa^* \leq \frac{\lambda_{max}\left((I_n+M)^{-1}\right)}{\lambda_{min}\left((I_n+M)^{-1}\right)} &= \frac{1+\lambda_{max}\left(M\right)}{1+\lambda_{min}\left(M\right)} \leq 1+\lambda_{max}\left(M\right),
\end{aligned}
\]
and therefore
\begin{equation}\label{eq:upper_bound_rho_PCG}
\rho_{EC}(\PGS,\pi_1)\leq 1-\frac{1}{n}\left[\frac{1+ \lambda_{min}(M)}{1+ \lambda_{max}(M)}\right]\,,
\end{equation}
which implies the other two inequalities in \eqref{eq:mixing_times_GS}.
\end{proof}

\subsection{Proof of Proposition \ref{prop:lower_bound_KL}}\label{sec:proof_lower_bound}

We need some preliminary lemmas.
\begin{lemma}\label{lem:var_log_conc_lower}
Let $X\sim \pi(x) \, \propto \, \exp(-U(x))$, with $U:\R\to\R$ convex and twice continuously differentiable, and $x_*=\arg\min_x U(x)$. Let also $a<x_*<b$ and $U(a)=U(b)=U(x_*)+1$ Then:
\begin{enumerate}
\item[(i)]  $\text{Var}(X) \geq d'(b-a)^2$ 
for some universal constant $d'>0$.
\item[(ii)] If $a \geq 0$ or $b \leq 0$, then $\text{Var}(|X|) \geq d(b-a)^2$ 
for some universal constant $d>0$.
\item[(iii)] If $U''$ is monotone, then $b - a \geq \sqrt{2/U''(x_*)}$ and therefore $\text{Var}(X) \geq d/U''(x_*)$ for a universal constant $d>0$. If moreover $a \geq 0$ or $b \leq 0$, then $\text{Var}(|X|) \geq d'/U''(x_*)$ for a universal constant $d'>0$.
\end{enumerate}
\end{lemma}
\begin{proof}
\emph{Part (i).}
Assuming $U(x_*)=0$ without loss of generality (w.l.o.g.), we have 
$$
Z=\int_{\R}\exp(-U(x))dx
\leq
(b-a)
+
\int_{-\infty}^a\exp(-U(x))dx
+
\int_b^{\infty}\exp(-U(x))dx\,.
$$
Using $U(b)=1\geq 0$ and $0=U(x_*)\geq U(b)+U'(b)(x_*-b)$, we deduce
$U(x)\geq U'(b)(x-b)\geq \dfrac{U(b)}{b-x_*}(x-b)
=\dfrac{x-b}{b-x_*}$ for $x\geq b$ and thus 
$\int_b^{\infty}\exp(-U(x))dx\leq b-x_*$. By similar arguments $\int_{-\infty}^a\exp(-U(x))dx\leq x_*-a$.
We thus obtain $Z\leq 2(b-a)$ and
\begin{align}\label{eq:lowe_dens_1}
\pi(x)&\geq \frac{e^{-1}}{2(b-a)}
&x\in(a,b)\,.
\end{align}
Then, given $A=(a,b)\subseteq \R$ and $B=\{x\in\R\,:\,|x-\mu|\geq (b-a)/4\}\subseteq \R$ with $\mu\in\R$, we have 
\begin{align}\label{eq:lowe_dens_2}
|A\cap B|
=
|A|-|A\cap B^c|
\geq 
|A|-|B^c|
=
(b-a)-(b-a)/2=(b-a)/2\,.
\end{align}
Taking  $\mu=\E[X]$ and combining \eqref{eq:lowe_dens_1} and \eqref{eq:lowe_dens_2} we obtain
$$
\mathbb{P}\left(|X-\E[X]|\geq \frac{b-a}{4}\right)\geq 
\frac{e^{-1}}{2(b-a)}\frac{b-a}{2}
\geq \frac{e^{-1}}{4}
$$
which implies
$$
\text{Var}(X)
=\E[|X-\E[X]|^2]
\geq\frac{e^{-1}}{4}\frac{(b-a)^2}{4^2}\,,
$$
as desired.

\emph{Part (ii).}
Assume without loss of generality that $a \geq 0$. Then, given $A=(a,b)\subseteq \R$ and $C=\{x\in\R\,:\,\left\lvert|x|-\mu \right\rvert \geq (b-a)/8\}\subseteq \R$ with $\mu\in\R$, reasoning as in the previous point we have that 
\begin{align}\label{eq:lowe_dens_3}
|A\cap C|
\geq
|A|-|C^c|
=
(b-a)-(b-a)/2=(b-a)/2\,.
\end{align}
Notice that the density of $|X|$, denoted by $\nu$, is such that $\nu(x) \geq \pi(x)$ for every $x \geq 0$ and in particular for $X \in A$ by assumption. Taking  $\mu=\E[|X|]$ and combining \eqref{eq:lowe_dens_1} and \eqref{eq:lowe_dens_3} we obtain
$$
\mathbb{P}\left(\biggl \lvert |X|-\E[|X|] \biggr\rvert \geq  \frac{b-a}{8}\right)\geq \int_{A \cap C} \pi(x) \d x \geq  
\frac{e^{-1}}{2(b-a)}\frac{b-a}{2}
\geq \frac{e^{-1}}{4}
$$
which implies
$$
\text{Var}(X)
=\E\left[\left( |X|-\E[|X|] \right)^2\right]
\geq\frac{e^{-1}}{4}\frac{(b-a)^2}{8^2}\,,
$$
as desired.

\emph{Part (iii).}
Let $a<x_*<b$ be such that $U(a)=U(b)=U(x_*)+1$.
Note that $a$ and $b$ exist and are unique by convexity of $U$ and integrability of $\exp(-U)$.
Assuming $U''$ non-increasing w.l.o.g., we have $U''(x)\leq U''(x_*)$ for $x\geq x_*$ which, together with $U'(x_*)=0$, implies $U(x)\leq U(x_*)+U''(x_*)(x-x_*)^2/2$ for $x\geq x_*$
and thus $b\geq x_*+\sqrt{2/U''(x_*)}$.
Thus
$(b-a)\geq(b-x_*)\geq \sqrt{2/U''(x_*)}$ and the two conclusions follow by parts (i) and (ii).
\end{proof}

\begin{lemma}\label{lem:gaps_lower_2}
Let $U(x)=x^2/(2c)+nh(x)$ with $x\in\R$ and $h=-\log\Phi$ as in \eqref{eq:pot_def}. 
Then $U''$ is non-increasing and
\begin{enumerate}
\item[(i)] If $cn\geq 3$, then $1/c \leq U''(x_*)\leq 5\log(cn)/c$.
\item[(ii)] Let $a < x_*$ such that $U(a) = U(x_*) + 1$. If $cn\geq 8$, then $a \geq 0$.
\end{enumerate}
\end{lemma}
\begin{proof}
We know that $h''$, and thus $U''$, is decreasing by the representation in \eqref{eq:h_curv} and \cite[Corollary 4]{mailhot1988some}.
By \eqref{eq:h_curv_1} and $h'=-\phi/\Phi$, we have
$h''(x)=h'(x)^2-xh'(x)$ for all $x$.

\emph{Part(i)}.
Take first $c=1$.
Since $U'(x_*)=x_*+nh'(x_*)=0$, we deduce $h'(x_*)=-x_*/n$ and
$$
1 \leq U''(x_*)=1+nh''(x_*)
=
1+n\left(\frac{x_*^2}{n^2}+\frac{x_*^2}{n}\right)
=
1+x_*^2\left(1+\frac{1}{n}\right)
\leq
1+2x_*^2\,.$$
Let $\tilde{x}=\sqrt{2\log(n)}$. Since $\tilde{x}>0$, we have
$h'(\tilde{x})\geq -2\phi(\tilde{x})
=-\sqrt{2/\pi}\exp(-\tilde{x}^2/2)=-\sqrt{2/\pi}n^{-1}\geq
-n^{-1}$. Since $n\geq 3$ we also have $\tilde{x}>1$, and thus
$U'(\tilde{x})=\tilde{x}+nh'(\tilde{x})\geq \tilde{x}-1\geq 0=U'(x_*)$. Since $U'$ is increasing we deduce $x_*\leq \tilde{x}=\sqrt{2\log(n)}$ and thus $U''(x_*)
\leq
1+2x_*^2\leq 1+4\log(n)\leq 5\log(n)$.

For general $c>0$, write $cU(x)=x^2/2+cnh(x)$ and apply the result with $c=1$ and $n$ replaced by $nc$.

\emph{Part(ii)}.
It suffices to prove the result for $c=1$, reasoning as in point (i). Let $x \leq x_*$. Since $U''$ is non-increasing we have that
\[
U(x) \geq U(x_*) + U''(x_*)\frac{(x-x_*)^2}{2},
\]
which implies
\begin{equation}\label{lower_bound_a}
x_* - a \leq \sqrt{\frac{2}{U''(x_*)}} \leq \sqrt{2}.
\end{equation}
Moreover let $\tilde{x} = \sqrt{\log n}$. Since $n \geq 8$, we have that $\tilde{x} > 0$ and therefore
\[
U'(\tilde{x}) \leq \sqrt{\log n} - \sqrt{\frac{n}{2\pi}} < U'(x_*) = 0.
\]
Since $U'(x)$ is increasing we deduce $x_* \geq \tilde{x} = \sqrt{\log n}$ Combining this with \eqref{lower_bound_a} we have that
\[
a \geq x_* - \sqrt{2} \geq \sqrt{\log n} - \sqrt{2} \geq 0,
\]
since $n \geq 8$.
\end{proof}

\begin{lemma}\label{lem:gaps_lower}
Consider model \eqref{eq:model} with $p = 1$, $m = 0$, $Q_0^{-1} = c>0$ and $x_i = 1$ for every $i$. Then, if $y_i = 1$ for every $i$ or $y_i = 0$ for every $i$, we have that
\[
\text{Var}_\pi(\beta_1) \geq dc/\left(\log (cn)\right) \quad \text{and} \quad \text{Var}_\pi(|\beta_1|) \geq d'c/\left(\log (cn)\right)
\]
for every $n\geq 8/c$ and some universal constants $d >0$ and $d' > 0$.
\end{lemma}
\begin{proof}
Assume without loss of generality that $y_i = 1$ for every $i$. Then $\pi(\beta_1)\propto\exp(-U(\beta_1))$ with $U$ as in Lemma \ref{lem:gaps_lower_2} and the result follows combining Lemma \ref{lem:gaps_lower_2} with Lemma \ref{lem:var_log_conc_lower}(iii).
\end{proof}
 We are finally ready to prove Proposition \ref{prop:lower_bound_KL}.
\begin{proof}
Define $\PMG=\Pi_{z \to \beta}\Pi_{\beta \to z}$, using a notation analogous to Proposition \ref{prop: equivalence_AC_MG}. 
It is well-known \citep[Section 3.3]{roberts2001markov} that $\PMG$ is $\pi(\beta)$-reversible. 

\emph{Part (i)}.
Choosing $f(\beta) = \beta_1$ in \eqref{eq:gap} we obtain
\[
\gap(\PMG) \leq \frac{\E\left[\text{Var}_\pi(\beta_1 \mid z)\right]}{\text{Var}_\pi(\beta_1)}\,.
\]
By \eqref{eq:posterior_beta} we have $\text{Var}_\pi(\beta_1 \mid z) = 1/(1/c+n)$ for every $z$ and $y$. We thus obtain 
\begin{equation}\label{eq:bound_gap}
\gap(\PMG) \leq \frac{1}{(1/c+n)\text{Var}_\pi(\beta_1)} \leq \frac{\log(cn)}{dc(1/c+n)},
\end{equation}
where the last inequality follows from the lower bound on $\text{Var}_\pi(\beta_1)$ in Lemma \ref{lem:gaps_lower}. Therefore, combining Proposition \ref{prop: equivalence_AC_MG} with \eqref{eq:general_lower_bound_gap} we have that
\[
\rho_{EC}(\PDA, \pi) \geq \rho_{EC}(\PMG, \pi_2) \geq 1- \frac{2\log(cn)}{dc(1/c+n)}\,,
\]
as desired.

\emph{Part (ii)}.
The upper bound follows immediately from Theorem \ref{theo:mixing_times_chi} and $XQ_0^{-1}X^T = cn$.

As regards the lower bound, choose $\mu \in \sP(\R^n\times \R)$ such that $\mu_2 \in \sP(\R)$ has density $\mu_2(\beta_1) \propto |\beta_1|\pi_2(\beta_1)$. Then
\[
f_2(\beta_1) = \frac{\mu_2(\beta_1)}{\pi_2(\beta_1)} = \frac{|\beta_1|}{\int_\R |\beta|\pi_2(\beta) \, \d \beta}
\]
and
\begin{equation}\label{upper_bound_RN}
\begin{aligned}
    \frac{\int \int (f_2(\beta')-f_2(\beta))^2\PMG(\beta, \d \beta') \pi(\d \beta)}{{2\text{Var}_\pi(f_2)}} &= \frac{\E[\text{Var}_{\pi}\left(|\beta_1| \mid z \right)]}{\text{Var}_{\pi}\left(|\beta_1| \right)}\\
    &\leq \frac{\E[\text{Var}_{\pi}\left(\beta_1 \mid z \right)]}{\text{Var}_{\pi}\left(|\beta_1| \right)} \leq \frac{\log(cn)}{d'c(1/c+n)},
\end{aligned}
\end{equation}
where the last inequality follows from the lower bound on $\text{Var}_\pi(|\beta_1|)$ in Lemma \ref{lem:gaps_lower} and $\text{Var}_\pi(\beta_1 \mid z) = 1/(1/c+n)$.

Combining \eqref{upper_bound_RN} with Corollary $7$ in \cite{wu2022minimax} we have that
\[
\tau_{\mathrm{mix},2}(\epsilon, \mu, \PMG)
    \geq 
    \frac{d'}{2}\left(\frac{1+cn}{\log (cn)}\right)\log\left( \frac{\chi^2( \mu, \pi)}{\epsilon}\right),
\]
and the result follows by Lemma \ref{lm: equivalence_AC_MG_chisquare}.
\end{proof}

\subsection{Proof of Proposition \ref{prop:starting_distribution}}
\begin{proof}
By definition of $\mu$ and the chain rule we have $\KL(\mu,\pi)
=
\KL(\mu_2,\pi_2)
$. We thus study the latter.
By definition of $\mu_2$ and $\pi$, and Bayes Theorem, we have $\frac{\mu_2(\beta)}{\pi_2(\beta)}\leq \frac{1}{m(y)}$ for every $\beta \in \R^p$, where
\[
m(y)= \int_{\R^p}\mathbb{P}(y_1 = g(z_1), \dots, y_n = g(z_n)\mid \beta)p(\d \beta) =\int_{\R^n}\prod_{i = 1}^n\Phi(\text{sgn}(2y_i-1)\eta_i)p_\eta(\d \eta)
\]
is the marginal likelihood of $y$, $p(\beta) = N(\beta \mid (0,\dots,0)^T, Q_0^{-1})$ is the prior of $\beta$ and $p_\eta(\eta) = N(\eta \mid (0,\dots,0)^T, M)$ with $M=XQ_0^{-1}X^T$ and $\eta=(\eta_1,\dots,\eta_n)=X\beta$. 
Thus $\KL(\mu_2, \pi_2) \leq -\log\left(
m(y)\right)$.
Also,
\[
\begin{aligned}
m(y)\geq \int_{K}\prod_{i = 1}^n\Phi(\text{sgn}(2y_i-1)\eta_i)p_\eta(\d \eta)
\geq
\Phi(-1)^n
p_\eta(K),
\end{aligned}
\]
with $K = \{ \eta\in\R^n \mid \|\eta\|^2 \leq 1\}\subset \{ \eta\in\R^n \mid |\eta_i| \leq 1\hbox{ for all }i\}$.
By
$p_\eta=N((0,\dots,0)^T, M)$ we have $p_\eta(K)\geq\gamma(K)$ with $\gamma=N((0,\dots,0)^T, \lambda_{max}(M)I_n)$. Then, using $[-n^{-1/2},n^{-1/2}]^n\subset K $ we obtain
\begin{align*}
\gamma(K)\geq 
\gamma([-n^{-1/2},n^{-1/2}]^n)
=
(N([-n^{-1/2},n^{-1/2}]\mid 0, \lambda_{max}(M)))^n
=
a(\,1/\sqrt{n\lambda_{max}(M)}\,)^n,
\end{align*}
with 
$a(\epsilon):=
\Phi(\epsilon)-\Phi(-\epsilon)=
2(\Phi(\epsilon)-0.5)$ for $\epsilon>0$.
Using $\Phi(\epsilon)-\Phi(0)\geq (\Phi(1)-\Phi(0))\epsilon$ for $\epsilon\in(0,1)$ we deduce $\Phi(\epsilon)-0.5\geq \min\{(\Phi(1)-0.5)\epsilon,\Phi(1)-0.5\}$ for $\epsilon\in(0,\infty)$ and thus
\begin{align}
\Phi(\epsilon)-0.5
\geq    
\min\{(\Phi(1)-0.5)\epsilon,\Phi(1)-0.5\}
\geq\frac{1}{4}\min\{1,\epsilon\}
=\frac{1}{4\max\{1,\epsilon^{-1}\}}
\geq\frac{1}{4(1+\epsilon^{-2})}\,,
\end{align}
where we used $\Phi(1)-0.5\geq 1/4$.
The above implies 
$1/a(\epsilon)\leq 2(1+\epsilon^{-2})$
and thus
\begin{align*}
-\log\left(
m(y)\right)\leq 
n\log(\Phi(-1)^{-1})
-n\log(a(\,1/\sqrt{n\lambda_{max}(M)}\,))
\leq 
2n
+n\log(2(1+n\lambda_{max}(M)))
\,,
\end{align*}
where we used $\log(\Phi(-1)^{-1})\leq 2$.
\end{proof}

\section{Empirical comparison between $\PDA$ and $\PGS$}\label{sec:comparison_supp}

\begin{figure}[h]
\centering
\includegraphics[width=.48\textwidth]{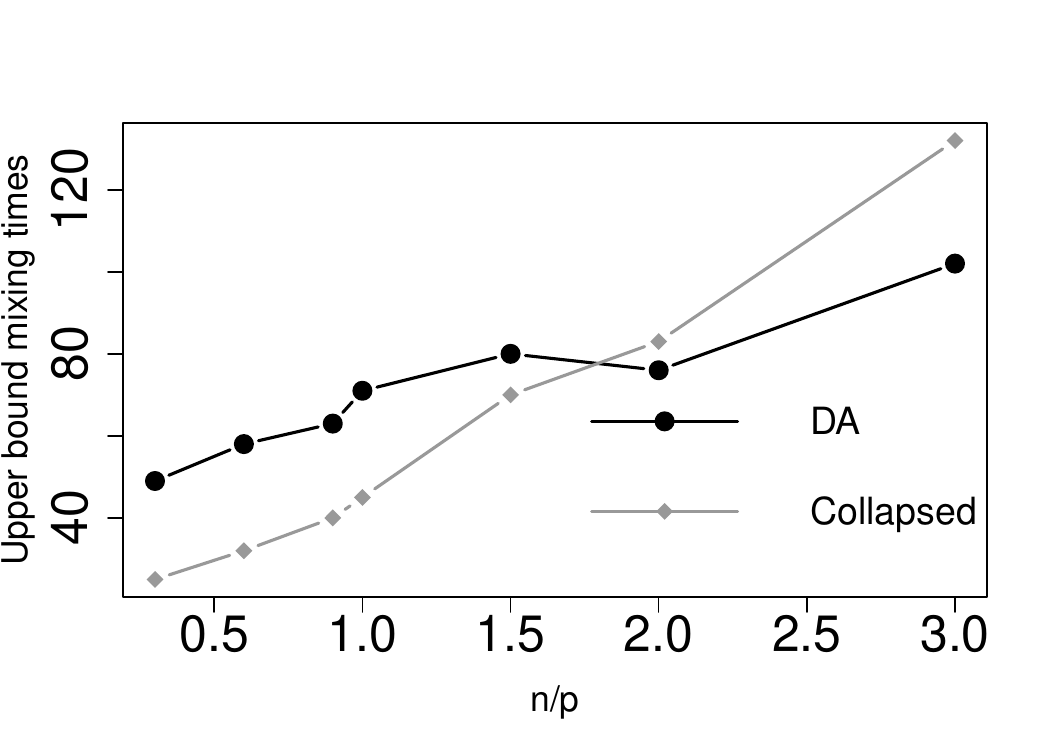} \quad
\includegraphics[width=.48\textwidth]{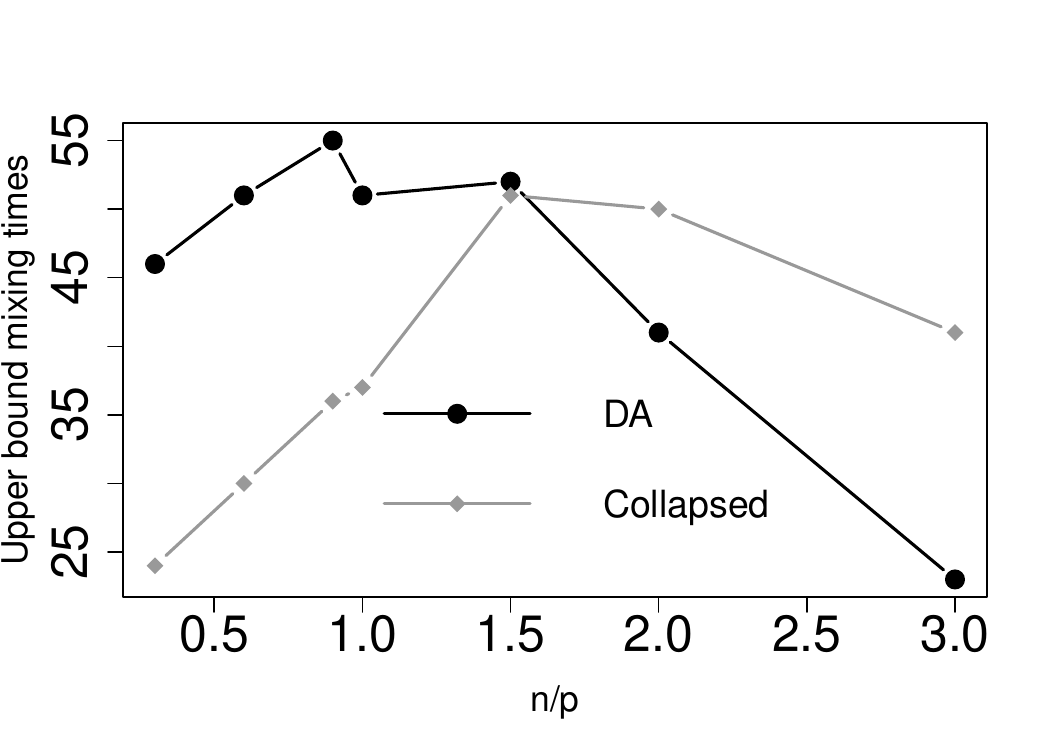}
 \caption{\small{
Upper bounds on 
$\tmix^{TV}(\epsilon, \mu,\PDA)$ and $\tmix^{TV}(\epsilon, \mu,\PGS)$ as a function $n/p$, with $p = 200$, $\epsilon = 0.1$, $Q_0^{-1} = 10 I_p$, $\mu(\d z, \d \beta) = N(\d\beta \mid 0, Q_0^{-1})\pi(\d z \mid \beta)$ and $X$ generated according to Assumption \ref{asmp:no_int} with $F = N(0,1)$. Bounds are obtained from \eqref{eq:bound_TV}, taking $L=200$ and estimating $\bar{d}(t)$ with $N = 500$ independent simulations of $\tau^{(L)}$. Observations are generated according to model \eqref{eq:probit_DA} (left) and with $y_i = 1$ (right).
  }}
 \label{fig:supp}
\end{figure}

In Figure \ref{fig:supp} the upper bounds on $\tmix^{TV}(\epsilon, \mu,\PDA)$ and $\tmix^{TV}(\epsilon, \mu,\PGS)$, with $\epsilon = 0.1$ and $X$ generated according to Assumption \ref{asmp:no_int}, are depicted as a function of the ratio $n/p$. Coherently with the results  of Corollary \ref{crl:noint}, since $c = 10$ is relatively large, when $n/p$ is small $\PGS$ yields faster convergence than $\PDA$ and viceversa when $n/p$ is large, even if the differences are overall moderate. 
Interestingly, the behaviour as $n$ grows differ depending on the data generating mechanism. The left plot of Figure \ref{fig:supp}, regarding data generated from the model, exhibits an increase of the mixing times; both chains seem instead to converge faster in the right plot, where $y_i = 1$ for every $i$. As mentioned in Section \ref{sec:discussion}, it would be interesting to complement the worst case results of Theorem \ref{theo:mixing_times} with upper bounds that depend on the observed $y$.

\section{Couplings}\label{sec_app:couplings}

\subsection{Auxiliary couplings}

We list below the auxiliary couplings for the algorithms used in Section \ref{sec:simulation}. Algorithms \ref{alg:max_reg} and \ref{alg:max_refl} contain the pseudocode for maximal couplings, i.e.\ that maximize the probability of the random variables being exactly equal after one step: Algorithm \ref{alg:max_refl} applies only to normal distributions with the same covariance matrix. Algorithms \ref{alg:it} and \ref{alg:crn} instead refer to optimal contractive couplings, minimizing the expected squared distance between the random variables after one step: Algorithm \ref{alg:it} only applies to univariate distributions, while Algorithm \ref{alg:crn} to Gaussian distributions. Finally, Algorithm \ref{alg:RwM_coupling} provides the pseudocode for a coupling between kernels of Random walk Metropolis operators starting from different points. See Sections $A$ and $B$ of the Supplementary Material of \cite{ceriani2024linear} for more details.

\begin{algorithm}
\begin{algorithmic}
 \State $\text{Sample }\mathbf{X} \sim p$
 \State $\text{Sample }W \sim U(0,1)$
 \If{$W p(\mathbf{X}) \le q(\mathbf{X})$}
 \State $\text{Set }  \mathbf{Y=X}$
 \Else
 \State Sample $  \mathbf{Y}^* \sim q$ and
 $W^* \sim U(0,1)$
 \While{$W^* q(\mathbf{Y}^*) < p(\mathbf{Y}^*)$}
 \State Sample $\mathbf{Y}^* \sim q$ and
 $W^* \sim U(0,1)$
 \EndWhile
  \State Set $\mathbf{Y=Y^*}$
 \EndIf
 \end{algorithmic}
 \caption{Maximal rejection coupling of $p,q \in \mathcal{P}(\mathbb{R}^d)$}
\label{alg:max_reg}
\end{algorithm}

\begin{algorithm}
\begin{algorithmic}
 \State Set $\mathbf{z}= \Sigma^{-1/2} (\boldsymbol{\xi}-\boldsymbol{\nu}), \, \mathbf{e}=\mathbf{z}/||\mathbf{z}||$ 
 \State Sample $\dot{\mathbf{X}} \sim N(\mathbf{0}, I), \, W \sim U(0,1)$
 \If{ $ W \le \exp\{-\frac{1}{2} \mathbf{z}^{\top}(2 \dot{\mathbf{X}} + \mathbf{z}) \}$} 
 \State Set $\dot{\mathbf{Y}} = \mathbf{\dot{X}}+\mathbf{z}$
 \Else
 \State Set $\mathbf{\dot{Y} = \dot{X}-2(e^\top \dot{X}) e}$
 \EndIf
 \State Set $\mathbf{X}= \Sigma^{1/2} \mathbf{\dot{X}}+ \boldsymbol{\xi}$
 \State Set $\mathbf{Y}= \Sigma^{1/2} \mathbf{\dot{Y}}+ \boldsymbol{\nu}$
 \end{algorithmic}
 \caption{Maximal reflection coupling of $N(\boldsymbol{\xi}, \Sigma)$ and $N(\boldsymbol{\nu}, \Sigma)$ }
\label{alg:max_refl}
\end{algorithm}

\begin{algorithm}
\begin{algorithmic}
	\State Sample $U \sim U(0,1)$
	\State Set $X = F_p^{-1}(U)$
	\State Set $Y = F_q^{-1}(U)$
    \end{algorithmic}
	\caption{Monotone map for $p, q \in \sP(\R)$ with distribution functions $F_p$ and $F_q$}
	\label{alg:it}
\end{algorithm}

\begin{algorithm}
\begin{algorithmic}
    \State Set $F = \Sigma^{1/2}$.
	\State Sample $\mathbf{Z} \sim N(\mathbf{0}, I)$
	\State Set $\mathbf{X} = \boldsymbol{\xi} + F\mathbf{Z}$
	\State Set $\mathbf{Y} = \boldsymbol{\nu} + F\mathbf{Z}$
    \end{algorithmic}
	\caption{Common random number coupling of $N(\boldsymbol{\xi}, \Sigma)$ and $N(\boldsymbol{\nu}, \Sigma)$ }
	\label{alg:crn}
\end{algorithm}

\begin{algorithm}
\begin{algorithmic}
	\State Sample $U \sim U(0,1)$
    \State Sample $(X', Y')$ with Algorithm \ref{alg:max_refl} such that $X' \sim N(x, \sigma^2)$ and $Y' \sim N(y, \sigma^2)$.
    \If{$U \leq p(X')/p(x)$}
    \State Set $X = X'$.
    \Else
    \State Set $X = x$.
    \EndIf
	\If{$U \leq q(Y')/q(y)$}
    \State Set $Y = Y'$.
    \Else
    \State Set $Y = y$.
    \EndIf
    \end{algorithmic}
	\caption{Coupling of Random walk Metropolis kernels $P(x, \cdot)$ and $P(y, \cdot)$ with proposal variance $\sigma^2$, which are invariant with respect to $p, q \in \sP(\R)$}
	\label{alg:RwM_coupling}
\end{algorithm}

\subsection{Couplings used in the illustrations}

Algorithms \ref{alg:meeting_time_PDA}, \ref{alg:meeting_time_PGS} and \ref{alg:meeting_time_PDA2} show the pseudocode for the couplings used in Section \ref{sec:simulation} to upper bound distance from stationarity through \eqref{eq:bound_TV}. We also use the notation $\theta^{(t)} = (z^{(t)}, \beta^{(t)})$ and $d(x, y) = \sqrt{\sum_{j = 1}^J(x_j-y_j)^2}$ for every $x, y \in \R^J$. In the pseudocode we implicitly assume that the couplings are relative to the corresponding steps in Algorithms \ref{alg:PDA}, \ref{alg:PGS} and \ref{alg:PDA2} respectively. Moreover we explored a range of values for $\epsilon$ and we obtained the tighter bounds with $\epsilon = 1/10$ (Algorithms \ref{alg:meeting_time_PDA} and \ref{alg:meeting_time_PDA2}) and $\epsilon = 1/1000$ (Algorithm \ref{alg:meeting_time_PGS}).

\begin{algorithm}[htbp]
\begin{algorithmic}
\State Set $\epsilon = 1/10$. Initialize $\theta^{(0)}_2 \sim \mu$ and $\theta^{(0)}_1 \sim \mu$, $\theta^{(t)}_1\mid \theta^{(t-1)}_{1} \sim \PDA(\theta^{(t-1)}_{1}, \cdot)$ for $t = 1, \dots, L$. 
\For{$t > L$}
    \If{$d\left(\theta^{(t-1)}_{1},\theta^{(t-L-1)}_{2}\right) > \epsilon$}
    \State Sample $\left(z^{(t)}_1, z^{(t-L)}_2\right)$ by coupling $\pi\left(  z_{1i} \mid \beta_1^{(t-1)}\right)$ and $\pi\left(  z_{2i} \mid \beta_2^{(t-L-1)}\right)$ 
    \State $\qquad$ for every $i = 1, \dots, n$ with Algorithm \ref{alg:it}.
    \State Sample $\left(\beta^{(t)}_1, \beta^{(t-L)}_2\right)$ by coupling $\pi\left(  \beta_{1} \mid z_1^{(t)}\right)$ and $\pi\left(  \beta_2 \mid z_2^{(t-L)}\right)$
    \State $\qquad$ with Algorithm \ref{alg:crn}.
    \Else
    \State Sample $\left(z^{(t)}_1, z^{(t-L)}_2\right)$ by coupling $\pi\left(  z_{1i} \mid \beta_1^{(t-1)}\right)$ and $\pi\left(  z_{2i} \mid \beta_2^{(t-L-1)}\right)$ 
    \State $\qquad$ for every $i = 1, \dots, n$ with Algorithm \ref{alg:max_reg}.
    \State Sample $\left(\beta^{(t)}_1, \beta^{(t-L)}_2\right)$ by coupling $\pi\left(  \beta_{1} \mid z_1^{(t)}\right)$ and $\pi\left(  \beta_2 \mid z_2^{(t-L)}\right)$
    \State $\qquad$ with Algorithm \ref{alg:max_refl}.
    \EndIf
    \State If $\theta^{(t)}_1 = \theta^{(t-L)}_{2}$, then return $\tau^{(L)} = t$.
\EndFor
\end{algorithmic}
\caption{(Sampling meeting times $\tau^{(L)}$ for $\PDA$)
\label{alg:meeting_time_PDA}}
\end{algorithm}

\begin{algorithm}[htbp]
\begin{algorithmic}
\State Set $\epsilon = 1/1000$. Initialize $z^{(0)}_2 \sim \mu$ and $z^{(0)}_1 \sim \mu$, $z^{(t)}_1\mid z^{(t-1)}_{1} \sim \PGS^n(z^{(t-1)}_{1}, \cdot)$ for $t = 1, \dots, L$. 
\For{$t > L$}
    \State Sample $(I_1, \dots, I_n) \simiid \text{Unif}(\{1, \dots, n\})$. 
    \If{$d\left(z^{(t-1)},z^{(t-L-1)}\right) > \epsilon$}
    \For{$i = 1, \dots, n$}
    \State Sample $\left(z^{(t)}_{1I_i}, z^{(t-L)}_{2I_i}\right)$ by coupling $\pi\left(  z_{1I_i} \mid z_{1, -I_i}\right)$ and $\pi\left(  z_{2i} \mid z_{2, -I_i}\right)$
    \State \qquad with Algorithm \ref{alg:it}.
    \EndFor
    \Else
    \For{$i = 1, \dots, n$}
    \State Sample $\left(z^{(t)}_{1I_i}, z^{(t-L)}_{2I_i}\right)$ by coupling $\pi\left(  z_{1I_i} \mid z_{1, -I_i}\right)$ and $\pi\left(  z_{2i} \mid z_{2, -I_i}\right)$
    \State \qquad with Algorithm \ref{alg:max_reg}.
    \EndFor
    \EndIf
    \State If $z^{(t)}_1 = z^{(t-L)}_{2}$, then return $\tau^{(L)} = t$.
    \EndFor
\end{algorithmic}
\caption{(Sampling meeting times $\tau^{(L)}$ for $\PGS^n$)
\label{alg:meeting_time_PGS}}
\end{algorithm}

\begin{algorithm}[htbp]
\begin{algorithmic}
\State Set $\epsilon = 1/10$. Initialize $\theta^{(0)}_2 \sim \mu$ and $\theta^{(0)}_1 \sim \mu$, $\theta^{(t)}_1\mid \theta^{(t-1)}_{1} \sim \PDA(\theta^{(t-1)}_{1}, \cdot)$ for $t = 1, \dots, L$. 
\For{$t > L$}
    \If{$d\left(\theta^{(t-1)}_{1},\theta^{(t-L-1)}_{2}\right) > \epsilon$}
    \State Sample $\left(\tilde{\beta}_{11}, \tilde{\beta}_{21}\right)$ according to Algorithm \ref{alg:RwM_coupling} \State $\qquad$ with $p = \pi\left(\beta_{11} \mid \beta_{1, -1}^{(t-1)}\right)$ and $q = \pi\left(\beta_{21} \mid \beta_{2, -1}^{(t-1)}\right)$.
    \State Sample $\left(z^{(t)}_1, z^{(t-L)}_2\right)$ by coupling $\pi\left(  z_{1i} \mid \tilde{\beta}_{11}, \beta_{1,-1}^{(t-1)}\right)$ and $\pi\left(  z_{2i} \mid \tilde{\beta}_{21}, \beta_{2,-1}^{(t-1)}\right)$ 
    \State $\qquad$ for every $i = 1, \dots, n$ with Algorithm \ref{alg:it}.
    \State Sample $\left(\beta^{(t)}_1, \beta^{(t-L)}_2\right)$ by coupling $\pi\left(  \beta_{1} \mid z_1^{(t)}\right)$ and $\pi\left(  \beta_2 \mid z_2^{(t-L)}\right)$
    \State $\qquad$ with Algorithm \ref{alg:crn}.
    \Else
    \State Sample $\left(\tilde{\beta}_{11}, \tilde{\beta}_{21}\right)$ according to Algorithm \ref{alg:RwM_coupling} \State $\qquad$ with $p = \pi\left(\beta_{11} \mid \beta_{1, -1}^{(t-1)}\right)$ and $q = \pi\left(\beta_{21} \mid \beta_{2, -1}^{(t-1)}\right)$.
    \State Sample $\left(z^{(t)}_1, z^{(t-L)}_2\right)$ by coupling $\pi\left(  z_{1i} \mid \tilde{\beta}_{11}, \beta_{1,-1}^{(t-1)}\right)$ and $\pi\left(  z_{2i} \mid \tilde{\beta}_{21}, \beta_{2,-1}^{(t-1)}\right)$ 
    \State $\qquad$ for every $i = 1, \dots, n$ with Algorithm \ref{alg:max_reg}.
    \State Sample $\left(\beta^{(t)}_1, \beta^{(t-L)}_2\right)$ by coupling $\pi\left(  \beta_{1} \mid z_1^{(t)}\right)$ and $\pi\left(  \beta_2 \mid z_2^{(t-L)}\right)$
    \State $\qquad$ with Algorithm \ref{alg:max_refl}.
    \EndIf
    \State If $\theta^{(t)}_1 = \theta^{(t-L)}_{2}$, then return $\tau^{(L)} = t$.
\EndFor
\end{algorithmic}
\caption{(Sampling meeting times $\tau^{(L)}$ for $\PDAmod$)
\label{alg:meeting_time_PDA2}}
\end{algorithm}

\end{document}